\documentclass[reprint,amsmath,amssymb,jap,aip,a4paper,%
               superscriptaddress,citeautoscript]{revtex4-1}

\usepackage{amssymb,amsmath,amsfonts}
\usepackage[utf8x]{inputenc}
\usepackage{graphicx,color}%
\usepackage{dcolumn}%
\usepackage{bm}%
\usepackage{placeins}
\usepackage{units} %
\usepackage{multirow}
\usepackage{pdfpages}
\usepackage{hyperref}
\hypersetup{
  pdfauthor={Koch, Granberg, Brink, Utt, Albe, Djurabekova, Nordlund},
  pdftitle=Local segregation versus irradiation effects in high-entropy alloys: Steady-state conditions in a driven system,
  pdfborder=0 0 0,
  colorlinks=true,
  urlcolor=blue,
  linkcolor=blue,
  citecolor=blue
}
\makeatletter
\AtBeginDocument{\let\LS@rot\@undefined}
\makeatother

\begin{document}
\frenchspacing
\raggedbottom

\title{Local segregation versus irradiation effects in high-entropy alloys:\\Steady-state conditions in a driven system}

\author{Leonie Koch}
\email{koch@mm.tu-darmstadt.de}
\affiliation{Fachgebiet Materialmodellierung, Institut f{\"u}r
  Materialwissenschaft, TU Darmstadt, Jovanka-Bontschits-Stra\ss{}e~2,
  D-64287 Darmstadt, Germany}

\author{Fredric Granberg}
\affiliation{Department of Physics, P.O.\ Box 43, FIN-00014 University
  of Helsinki, Finland}

\author{Tobias Brink}
\affiliation{Fachgebiet Materialmodellierung, Institut f{\"u}r
  Materialwissenschaft, TU Darmstadt, Jovanka-Bontschits-Stra\ss{}e~2,
  D-64287 Darmstadt, Germany}

\author{Daniel Utt}
\affiliation{Fachgebiet Materialmodellierung, Institut f{\"u}r
  Materialwissenschaft, TU Darmstadt, Jovanka-Bontschits-Stra\ss{}e~2,
  D-64287 Darmstadt, Germany}

\author{Karsten Albe}
\affiliation{Fachgebiet Materialmodellierung, Institut f{\"u}r
  Materialwissenschaft, TU Darmstadt, Jovanka-Bontschits-Stra\ss{}e~2,
  D-64287 Darmstadt, Germany}

\author{Flyura Djurabekova}
\affiliation{Department of Physics, P.O.\ Box 43, FIN-00014 University
  of Helsinki, Finland}
\affiliation{Helsinki Institute of Physics, P.O.\ Box 43, FIN-00014 University
  of Helsinki, Finland}

\author{Kai~Nordlund}
\affiliation{Department of Physics, P.O.\ Box 43, FIN-00014 University
  of Helsinki, Finland}

\date{11 September 2017}

\begin{abstract}
We study order transitions and defect formation in a model high-entropy alloy (CuNiCoFe) under ion irradiation by means of molecular dynamics simulations.
Using a hybrid Monte-Carlo/molecular dynamics scheme a model alloy is generated which is thermodynamically stabilized by configurational entropy at elevated temperatures, but partly decomposes at lower temperatures by copper precipation. Both the high-entropy and the multiphase sample are then subjected to simulated particle irradiation. The damage accumulation is analyzed and compared to an elemental Ni reference system.
The results reveal that the high-entropy alloy---independent of the initial configuration---installs a certain fraction of short-range order even under particle irradiation. Moreover, the results provide evidence that defect accumulation is reduced in the high-entropy alloy. This is because the reduced mobility of point defects leads to a steady state of defect creation and annihilation. The lattice defects generated by irradiation are shown to act as sinks for Cu segregation.\\[-\baselineskip]
  \begin{center}
    \rule{10cm}{0.4pt}
  \end{center}
  \noindent
  \footnotesize This article may be downloaded for personal use
  only. Any other use requires prior permission of the author and AIP
  Publishing. The following article appeared in
  \href{https://doi.org/10.1063/1.4990950}{J.\ Appl.\ Phys.\
    \textbf{122}, 105106 (2017)} and may be found at
  \url{https://doi.org/10.1063/1.4990950}.
\end{abstract}

\maketitle

\section{Introduction}
\label{sec1}

High-entropy alloys (HEAs) constitute a relatively new class of materials, which have recently attracted considerable attention in the field of high-performance materials \cite{WanGuoLiu}.
They consist of at least four to five principal elements occuring in an equimolar or near equimolar ratio, such that their fractions do not drop below
\unit[5]{$\%$} or exceed \unit[35]{$\%$} \cite{2016-Ye,Otto2013, 2016-Pickering}.
HEAs owe their name mainly to the large contribution of configurational entropy to the Gibbs free energy. It is supposed that the influence of the entropy stabilizes random solid solutions even at lower temperatures \cite{TasDenPra, Yeh}.
However, mixing enthalpies and the atomic size mismatches between the components decisively contribute to the phase selection criterion at lower temperatures \cite{2016-Pickering} and the formation of secondary phases can often not be completely avoided \cite{WidHuhMai13}.

The interest in high-entropy alloys is mostly triggered by their outstanding mechanical properties, making them alternatives for certain superalloys or metallic
glasses \cite{2016-Ye, 2016-Pickering, TsaiWei,LiuWuHe,XieBraTho14, TiaVarChe13, KaoYehChi08, HagKoeCat14, TiaDelChe13, GuoHuNg13, NgGuoLua13}. However, high-entropy alloys are not only characterized by
a large configurational entropy,  but also by high atomic-level stresses arising from local lattice distortions due to atomic size differences \cite{Egami2014}. 
It has been reported that  structural and chemical disorder affect both defect kinetics and heat dissipation, which is of particular interest in the context of radiation resistant materials~\cite{2015-Zhang}. Recent molecular dynamics (MD) simulations revealed that defect concentrations can be significantly reduced during irradiation in equiatomic multicomponent alloys as compared to elemental metals \cite{2016-Granberg,2016b-Granberg}. This advantageous robustness 
against radiation damage can be ascribed to a reduced defect mobility and consequently a smaller growth rate of defect clusters \cite{2016-Granberg,2016b-Granberg,Yang2016}.
Additionally, chemical disorder decreases thermal conductivity, which is also assumed to prevent the formation of large defect clusters by facilitating defect annihilation \cite{Egami2014, 2015-Zhang, 2016-Ullah}. 
If, on the other hand, a HEA has a tendency towards multiphase formation at lower temperatures,
particle irradiation can also drive the system into thermodynamic equilibrium and lead to precipitiation \cite{Yang2016}.
Thus, in irradiated high-entropy alloys there is a delicate interplay of thermodynamic driving forces due to configurational entropy, mixing/demixing tendencies, and irradiation-induced far-from-equilibrium conditions.

The objective of the present study is to address this issue in a model four-component CuNiCoFe alloy in thermodynamic equilibrium and under particle irradiation, as well as to study the transition from a single phase HEA to multiphase or so-called  compositionally complex alloy.
Using a hybrid simulation scheme consisting of alternating Monte-Carlo (MC) and MD steps, we start by generating a model alloy which is truly stabilized by configurational entropy at elevated temperatures. We then show that this structure decomposes at lower temperatures by precipation of small copper clusters. Both the high-entropy and the partly decomposed sample are then subjected to a series of 1,500 recoil events and the damage accumulation is analyzed and compared to elemental Ni as reference system. Moreover, we show that segregation effects occur at irradiation-induced defects, promoting the formation of precipitates.

\section{Methods}
\label{sec2}

\subsection{Interatomic potential}

We chose a CuNiCoFe alloy as a model system for HEAs with a tendency for Cu segregation at low temperatures. The embedded atom method \cite{DawBas} (EAM) was used to describe interatomic interactions in the multicomponent system. Parametrizations for the elemental interactions were taken from Zhou \textit{et al} \cite{Misfit}. The missing cross terms were created using the potential generator from the same authors, which is based on a single element mixture procedure \cite{Misfit}.
Because of the ferromagnetic elements, magnetism and magnetic transitions could influence the evolution of chemical ordering. EAM potentials cannot explicitly capture these effects and include the magnetic energy contributions only implicitly. For this reason, our system can only serve as a model for nonmagnetic HEAs or HEAs without magnetic transitions. In order to nonetheless ensure the reliability of the potential, we validated it against reference data in Appendix~\ref{sec:pot-val}.

\subsection{Equilibrium simulations}

Initially, atoms were distributed randomly on a regular FCC lattice of about $N = 100,000$ sites, such that an equimolar configuration was obtained. We applied 3-dimensional periodic boundary conditions and the initial lattice constant was approximated according to Vegard's law \cite{Jac}. After the construction of the initial structure, we minimized its free energy with a mixed MC/MD procedure, carried out using the open source code \textsc{lammps} \cite{Pli95}. For the MC steps, we used the variance-constrained semi-grand-canonical (VC-SGC) ensemble \cite{SadErhStu12}. This ensemble is based on the semi-grand-canonical ensemble with an added constraint on the overall composition. Thus, it resembles a canonical ensemble, but allows Gaussian variations of the concentration, the extent of which is controlled by the parameter $\kappa$. We chose a value of $\kappa = 10^3$ in our simulations. The VC-SGC ensemble allows simulations inside miscibility gaps and has a performance advantage over the canonical ensemble, which cannot be parallelized easily \cite{SadErhStu12}. The miscibility is controlled via the chemical potential differences. The chemical potential differences with regard to copper were determined as $\Delta \mu_\text{Cu--Ni} = \unit[0.9]{eV}$, $\Delta \mu_\text{Cu--Co} =\unit[0.85]{eV}$, and $\Delta \mu_\text{Cu--Fe}= \unit[0.7]{eV}$  as described in Ref.~\onlinecite{Brink2016a}. The values at hand lead to a miscible CuNiCoFe system at \unit[800]{K}.

The MC/MD simulations were performed at different target temperatures $T = \unit[800]{K}, \unit[750]{K}, \ldots, \unit[400]{K}$, with $N/4$ MC trial moves followed by 20 MD steps. For the MD steps, we used a timestep of \unit[1]{fs}, as well as a Nos\'e--Hover thermostat and a Parinello--Rahman barostat at temperature $T$ and ambient pressure. In total, the simulations were run for $1,000,000$ MD steps (including $50,000$ MC cycles), after which the potential energy was definitely converged.

\subsection{Irradiation simulations}

We simulated irradiation of the structures using the approach
introduced in Ref.~\onlinecite{2016-Granberg} with the \textsc{parcas} MD code~\cite{parcas}.
Simulations were started from
either pure Ni, a CuNiCoFe cell with random element distribution, or the equilibrated sample from the MC/MD run at \unit[400]{K}. After
equilibration at room temperature to zero pressure, the simulation
cell size was fixed and a series of subsequent \unit[5]{keV} recoils was
initiated in the cell. The recoils were performed by assigning a velocity vector with a magnitude corresponding to the recoil energy and a random direction to the atom closest to the center of the simulation cell. Each recoil was simulated for \unit[30]{ps}, which was
sufficient to cool the cell back down to ambient
temperature by Berendsen temperature scaling \cite{Ber84} in a thin layer at the simulation box boundaries. After this, all atom coordinates
were shifted by a displacement vector with components randomly selected in the
interval $[0,L_d]$, where $L_d$ is the cell size in each of the
3 dimensions $d$. After the shift, atoms outside the boundaries
were wrapped back into the cell according to the periodic boundary conditions.
This procedure ensures homogeneous irradiation of the entire simulation
cell. Experimentally, this is comparable, e.g., to prolonged neutron or high-energy ion irradiation of a segment inside the material.

The irradiation simulations used the same interatomic potential as
the MC/MD simulations, except that for small interatomic separations
(well below the equilibrium nearest-neighbor distance) the potentials
are smoothly joint to the universal repulsive
Ziegler--Biersack--Littmark (ZBL) interatomic potential \cite{ZBL}.  The
ZBL-96 electronic stopping \cite{SRIM96} was applied on all
atoms with kinetic energies higher than \unit[1]{eV}.  Monitoring of the
simulations showed that about \unit[20]{$\%$} of the initial recoil energy was
lost to electronic stopping, i.e., the nuclear deposited damage
energy per recoil is
about \unit[4]{keV}. The total deposited nuclear energy was extracted by summing up the damage energies and then used to calculate the
displacements-per-atom (dpa) value, a standard unit for
radiation damage exposure \cite{ASTMFe,OECDdpareport}.
This irradiation simulation approach has previously
been demonstrated to lead to a good agreement with experiments on other HEAs \cite{2016-Granberg,Zha16c}.

\subsection{Analysis of short-range order and lattice defects}

In order to quantify the chemical ordering, we used the Warren--Cowley short-range order (SRO) parameters $\alpha_1$ and $\alpha_2$ adapted to a multicomponent alloy \cite{1950-Cowley}
\begin{equation}
  \alpha_n^{ij} = 1-\frac{P_n^{ij}}{c_{j}},
\end{equation}
where the subscript $n \in \{1,2,\dots\}$ refers to the $n^{\text{th}}$ coordination sphere, $P_n^{ij}$ describes the conditional probability of an atom of type $j$ being adjacent to an atom of type $i$, and $c_{j}$ is the concentration of atom type $j$. A Warren--Cowley parameter of zero represents the case of an ideal solid solution with no tendency for clustering. Ordered structures (attractive SRO) can be assumed if $\alpha$ is positive, while negative values reflect preferences for segregation (repulsive SRO).  We calculated the SRO parameters for the equilibrium structures as a time average over 20 snapshots taken during 2000 simulation steps. Those snapshots were minimized prior to analysis to remove thermal fluctuations and to obtain the \unit[0]{K} ground-state values. For the irradiated cells, the SRO analysis was performed in the same manner, except that averaging was only performed where noted.

We used the open source application \textsc{ovito} \cite{Stu10} for the analysis and visualization of simulation results: Atomic volumes were calculated by Voronoi tesselation of minimized samples \cite{Voronoi1908, Voronoi1908a, Voronoi1909,
  Brostow1998}, defective atoms were marked by common neighbor analysis (CNA) with an adaptive cutoff \cite{Honeycutt1987, Stukowski2012}, and dislocation lines were identified using the dislocation extraction algorithm (DXA) \cite{DXA}. Additionally, we also performed a Wigner--Seitz cell defect analysis \cite{Nor97f}. This analysis constructs the space-filling
Wigner--Seitz cells of the perfect underlying FCC lattice, labeling
empty cells as vacancies and doubly-filled ones as
interstitials. Since this analysis is space-filling, it gives a
definite determination of whether a defect region is of vacancy or interstitial type (complex defects may involve both, but then the difference in the number of vacancies vs.\ interstitials of the same cluster determines its character \cite{Nor99}). The supplemental information provided by this analysis helps to identify point defects, which cannot be treated by DXA.

\section{Results and Discussion}
\label{sec3}

\subsection{CuNiCoFe in thermodynamic equilibrium}
\label{sec:res:equil}

\begin{figure}
  \centering
  \makebox[\linewidth]{%
    \includegraphics[]{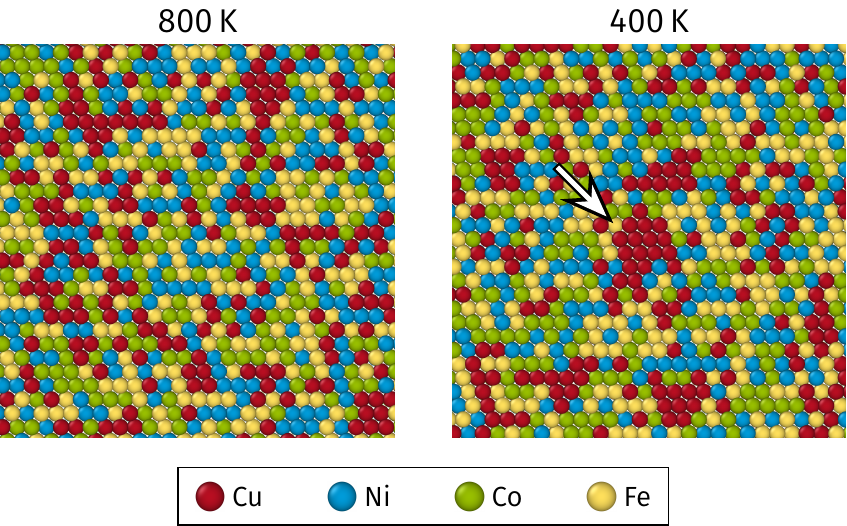}}
  \caption{Snapshots of the CuNiCoFe alloy, equilibrated in the VC-SGC ensemble at \unit[800]{K} (left) and \unit[400]{K} (right). The arrow highlights a clustering of copper atoms, indicating that phase separation occurs at lower temperatures.}
  \label{fig:VCSGC800KL400KR}
\end{figure}
\begin{figure}[t!]
  \centering
  \makebox[\linewidth]{%
    \includegraphics[]{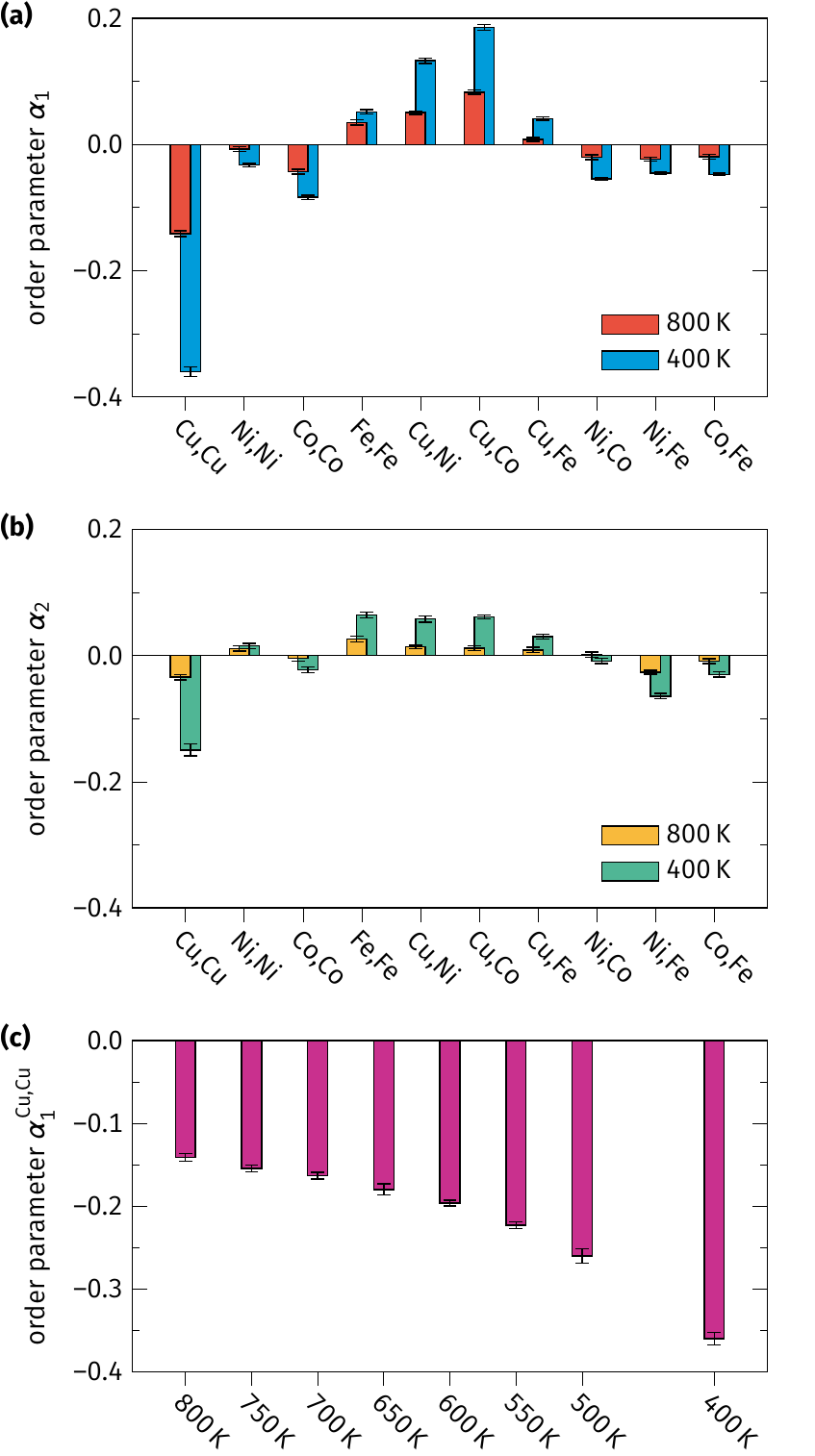}}
  \caption{Warren--Cowley parameters $\alpha_1$ (a) and $\alpha_2$ (b) for the
    CuNiCoFe alloy system at \unit[800]{K} and \unit[400]{K}. (c) The change of $\alpha_1^\text{Cu,Cu}$ with temperature. Increasing Cu--Cu ordering tendencies can be observed with decreasing temperature. All values are averaged over 20 snapshots at \unit[0]{K}, with the error bars representing the standard deviation.}
  \label{fig:WC_VCSGC}
\end{figure}
Figure \ref{fig:VCSGC800KL400KR} shows the final structures after equilibration in the VC-SGC ensemble at \unit[800]{K} (left) and at \unit[400]{K} (right). The configurations differ slightly in their atomic arrangement. Whereas Cu (red atom type) appears to be randomly distributed at \unit[800]{K}, it tends to form small clusters in the structure at \unit[400]{K}, indicated by the white arrow.
The Warren--Cowley parameters at both temperatures are shown in Fig.~\ref{fig:WC_VCSGC}.
All $\alpha_n$ values more or less deviate from the ideal solid solution in either the positive or negative direction. This behavior is most pronounced for Cu. On average we find a Warren--Cowley parameter for Cu--Cu of $-0.15$ for the first neighbor shell at \unit[800]{K}, indicating a slight tendency to form small Cu clusters (positive $\Delta H_\text{mix}$). Accordingly, all other $\alpha_1^{\text{Cu--}j}$ values are positive. Except for Fe--Fe
interactions, all remaining parameters are slightly negative, indicating miscibility of Ni, Co, and Fe.
The $\alpha_2$ parameter exhibits no strong indication for medium-range ordering at \unit[800]{K}.

If we repeat the analysis in the same sample equilibrated at \unit[400]{K}, we see that the ordering tendencies become more distinct and extend into the second neighbor shell.
Although the Warren--Cowley parameters for non-Cu pairs also change in magnitude, they still indicate an approximately random solid solution  of Co, Ni, and Fe without any remarkable clustering.
This can also be inferred from the snapshots in Figure \ref{fig:VCSGC800KL400KR}.
Figure~\ref{fig:WC_VCSGC}c further shows that the Cu segregation tendencies increase non-linearly with decreasing temperature. 
Note that we do not observe any structural phase transitions, such as a transition from an FCC to a BCC lattice or even amporphization.

These results may be interpreted as follows: The contribution of the configurational entropy at \unit[400]{K} is insufficient to compete against the enthalpy of mixing. Since the composition of the system is fixed, it will demix by locally forming copper-rich clusters.
The formation of two separate solid solution phases (Cu-rich and Cu-depleted) has already been confirmed experimentally by Otto \textit{et al.} \cite{Otto2013} The observed decomposition is more pronounced than in the present simulation, which may either be due to the different composition in the experiment (CoCrFeMnCu), or the higher solubility of Cu in the EAM potential (see appendix). The authors found that the Cu segregation is driven by positive binary mixing enthalpies between copper
and the remaining components \cite{Otto2013}.
Thus, when decreasing the temperature in experiment, there is a competition between the increasing thermodynamic driving force for decomposition and the decreasing diffusivity within the system. For this reason, HEAs often represent kinetically trapped metastable configurations at lower temperatures and local segregation occurs as an ageing phenomenon.

\begin{figure}
  \centering
  \includegraphics[]{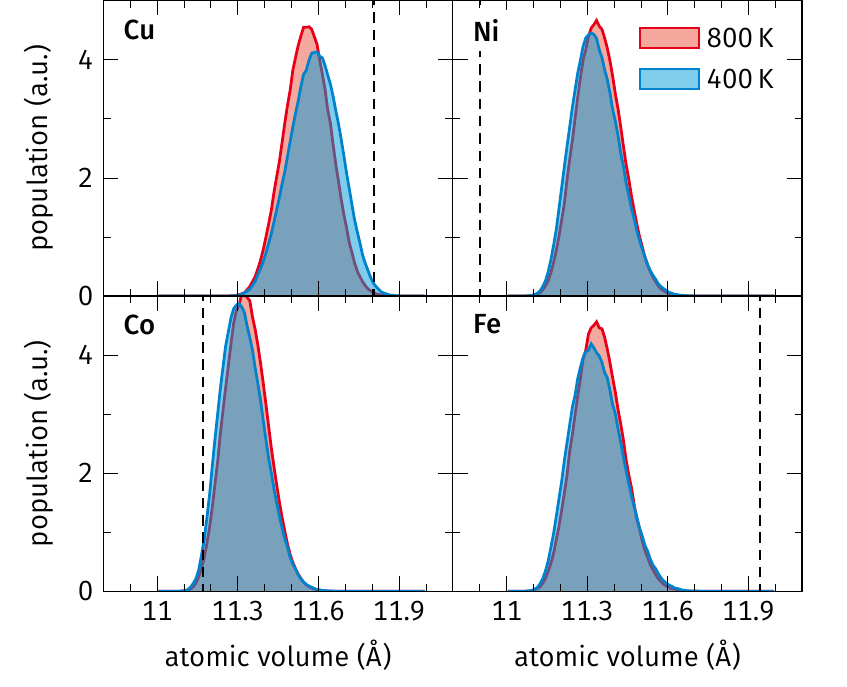}
  \caption{Histograms of the atomic volumes at \unit[0]{K} for the HEA equilibrated at \unit[400]{K} (red) and \unit[800]{K} (blue). The dashed lines indicate the ground state atomic volume of the single-element FCC structures. Atomic volumes are obtained from Voronoi tesselation\cite{Voronoi1908, Voronoi1908a, Voronoi1909, Brostow1998} of 20 snapshots at \unit[0]{K}.}
  \label{fig:ScatterPlotVoronoiVolType}
\end{figure}
When comparing the atomic volumes as a function of the atom type (see Fig.~\ref{fig:ScatterPlotVoronoiVolType}), it can be seen that Cu has slightly larger values than Co, Ni, or Fe. Both Ni and Co possess higher atomic volumes in the multicomponent system than in their elemental state. In contrast, Cu and Fe exhibit smaller atomic volumes compared to their pure structures. At \unit[400]{K}, the system starts demixing into a ternary Co, Ni, and Fe solid solution and into a copper-rich phase. The copper atoms expand slightly, while the ternary solid solution becomes slightly denser. This brings Cu, Co, and Ni closer to their preferred atomic volume. Only Fe deviates from this trend, but exhibits a positive Warren--Cowley parameter, indicating a preference for unlike neighbors. This changed chemical environment may influence the equilibrium volume.

In summary, these observations suggest that the small scale decomposition is driven largely by the mixing enthalpy at low temperatures, but additionally enhanced by atomic size effects.

\subsection{Irradiation}

\begin{figure*}
  \centering
  \makebox[\linewidth]{%
    \includegraphics[]{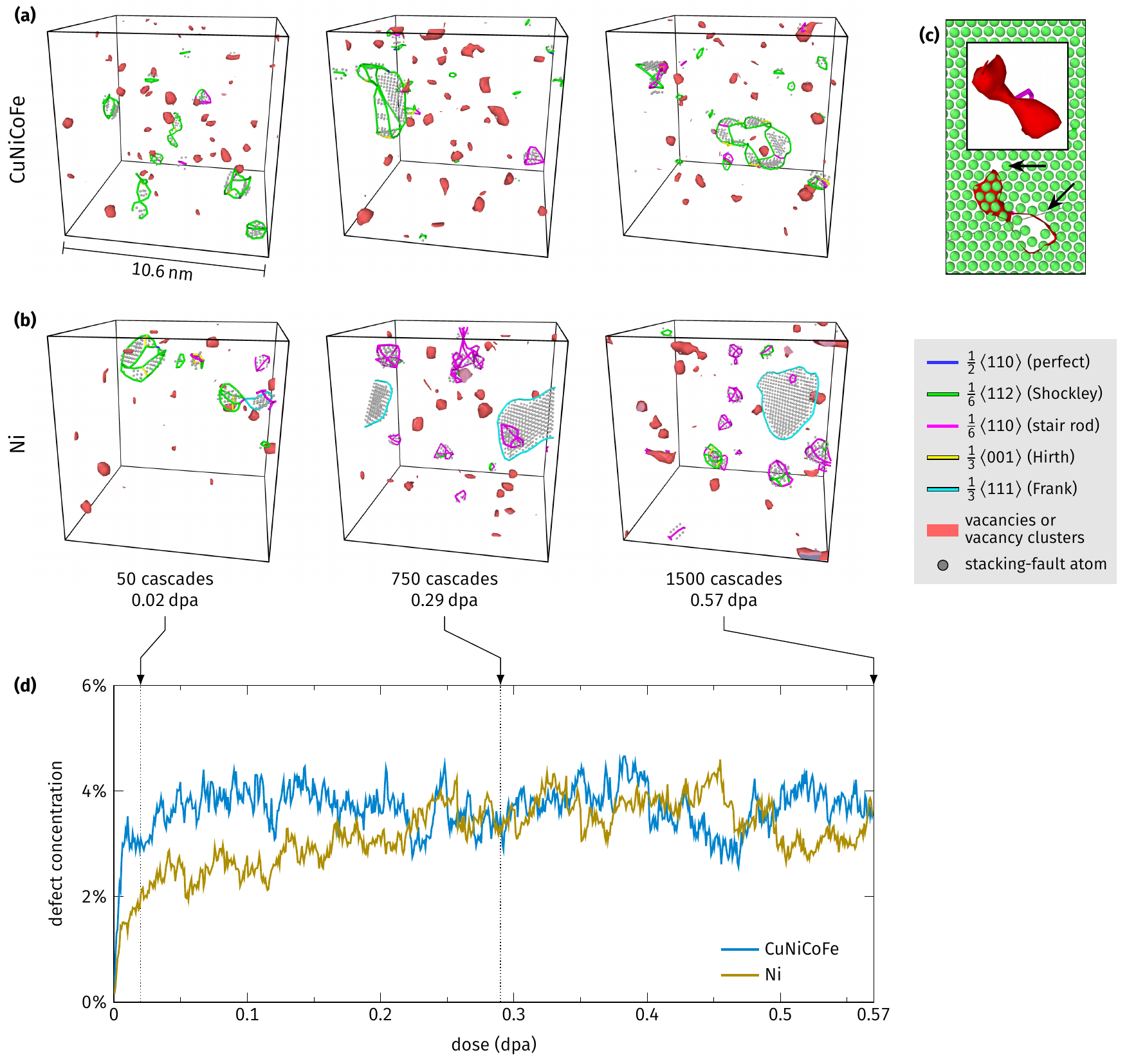}}
  \caption{Analysis of the build-up of lattice defects during
    irradiation. (a) DXA analysis of the initially segregated HEA at
    different irradiation doses. (b) The same for a Ni sample. Empty
    space represents the perfect FCC lattice; the structures do not
    collapse during irradiation. Green lines indicate $\langle 112
    \rangle$ partial dislocations, turquoise lines indicate a Frank
    loop, purple lines belong to stacking fault tetrahedra, and red
    surfaces enclose defects that cannot be recognized by DXA. Videos
    of these simulations can be found in the \hyperref[supplemental]{supplementary material}
    (\texttt{CuNiCoFe-ordered-DXA-during-irradiation.avi} and
    \texttt{Ni-DXA-during-irradiation.avi}). (c) A closer look at those
    red regions reveals that they represent vacancies and vacancy
    clusters. In (d), a plot of the concentration of defective atoms
    as identified by CNA is shown as a function of the irradiation
    dose. In agreement with the DXA results, we can see that the HEA
    quickly reaches a high defect concentration that saturates around
    \unit[4]{\%}. These defects consist mostly of vacancies and small
    dislocation networks. Pure Ni builds up the defect concentration
    more slowly. At first---similar to the HEA---vacancies and small
    dislocation networks appear, then these start disappearing in
    favor of stacking-fault tetrahedra and a Frank loop.}
  \label{fig:irrad-dxa}
\end{figure*}

\begin{figure}
  \centering
  \makebox[\linewidth]{%
    \includegraphics[]{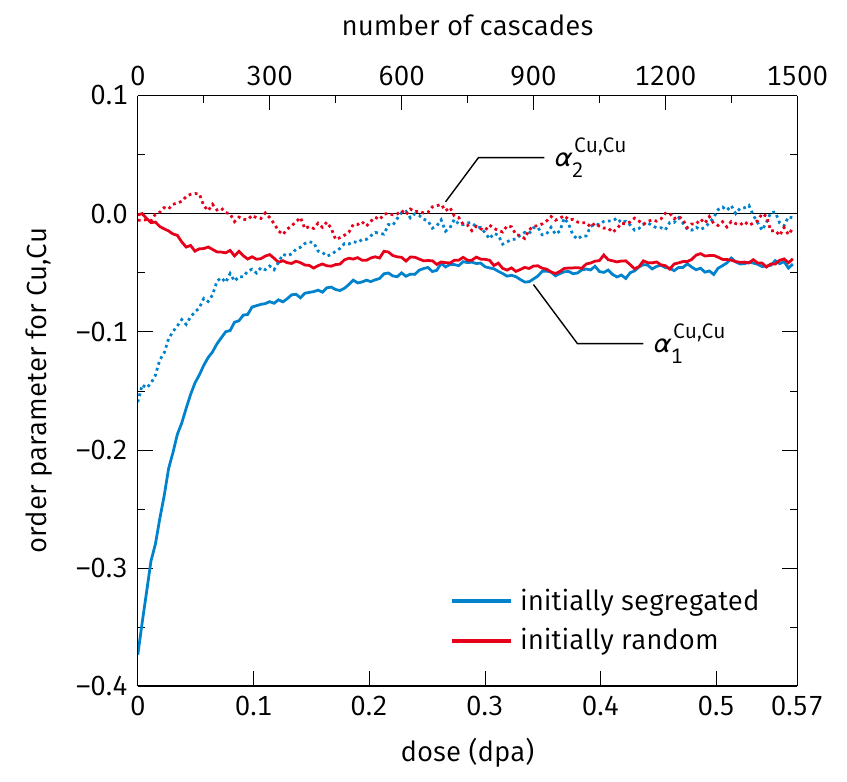}}
  \caption{Evolution of the SRO during irradiation. Only the Cu--Cu parameters are shown, the graphs for all parameters can be found in Fig.~S1 in the \hyperref[supplemental]{supplementary material}. The sample which was equilibrated using VC-SGC (blue) and the one which was initially random (red) converge to the same values for the order parameters.}
  \label{fig:irrad}
\end{figure}

\begin{figure}[t!]
  \centering
  \makebox[\linewidth]{%
    \includegraphics[]{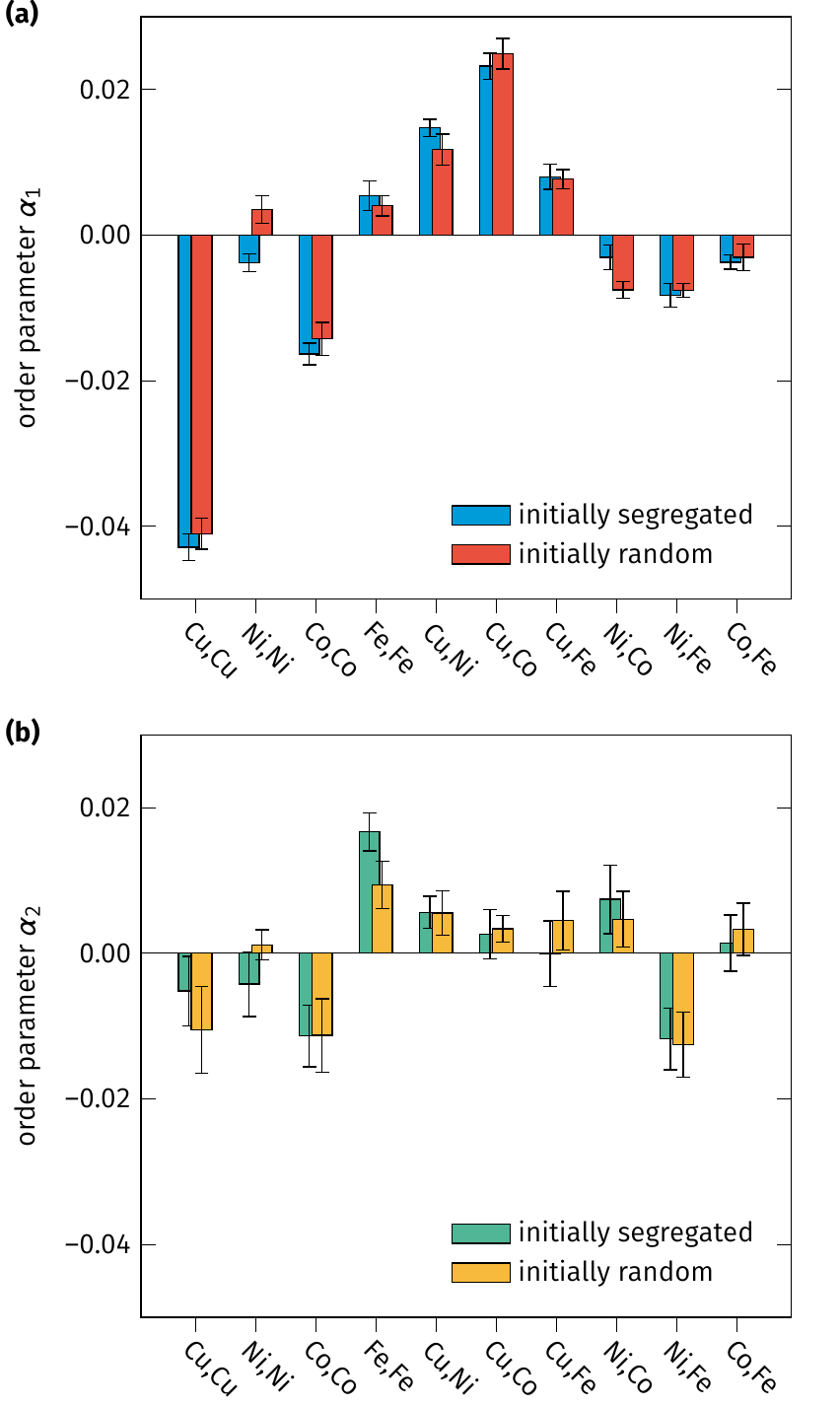}}
  \caption{Final order parameters for the first (a) and second neighbor shell (b) after irradiation with a dose of \unit[0.57]{dpa}. The values are averaged over the last 10 frames of the irradiation simulation, with the error bars representing the standard deviation.}
  \label{fig:irrad-final-state}
\end{figure}

Thus far, we studied the thermodynamic stability of a \mbox{CuNiCoFe} solid solution.  In section~\ref{sec:res:equil}, we demonstrated by means of hybrid MC/MD simulations that \mbox{CuNiCoFe} exhibits a tendency for decomposition at lower temperatures (400~K). These results suggest that in this case a random distribution is only stabilized by kinetic inhibition. In the following, we will therefore discuss the effect of cascade events on the evolution of the chemical order, focusing on the question, if the energy that is released during recoil events enables the decomposition of Cu. We irradiate both the equilibrated, partially segregated sample obtained by MC/MD simulation at \unit[400]{K} (``initially segregated''), and a sample with completely random element distribution (``initially random''). Additionally, we investigate the structural damage due to irradiation by comparison with a Ni sample.

\subsubsection{Lattice defects}

Earlier work comparing irradiation damage in elemental metals and disordered alloys suggests that the FCC structure is retained throughout irradiation with $1,500$ recoils \cite{2016-Granberg}. Figure~\ref{fig:irrad-dxa}(a) shows that this is also the case for the initially segregated HEA sample (equilibrated with MC/MD at \unit[400]{K}). Since the structural damage in the initially random HEA is similar, we omit this sample from the discussion at hand. For comparison, Figure~\ref{fig:irrad-dxa}(b) shows a pure Ni reference sample which underwent the same irradiation treatment. Most irradiation damage in these samples manifests itself by the formation of dislocation loops, stacking-fault tetrahedra,
and single vacancies or vacancy clusters. Vacancy clusters and extended defect agglomerates cannot be identified by the DXA and are indicated by red defect meshes. A visualization of atom positions in Fig.~\ref{fig:irrad-dxa}c reveals that these regions are dominantly characterized by vacant lattice sites.
The nature of these regions has also been confirmed by a Wigner--Seitz cell defect analysis, which shows that the positions of the distorted regions
generally coincide with vacancy-type defect clusters (see videos \verb|cascade-CuNiCoFe.avi| and \verb|cascade-Ni.avi| in the \hyperref[supplemental]{supplementary material}).
These vacancy-type defects are homogeneously distributed within the structure.

The irradiation process and therefore the formation and evolution of defects can be divided into three stages. Figures~\ref{fig:irrad-dxa}a and b visualize
one representative configuration for each of the three stages.  Figure~\ref{fig:irrad-dxa}d depicts the defect concentration as a function of the irradiation dose.
Initially, the HEA shows a higher defect accumulation than the pure Ni system, which could
be caused by higher atomic level stresses within the multicomponent system. This results in a reduced defect formation energy and thus increased number of defects\cite{Egami2014}. In turn, pure Ni exhibits a higher recombination
rate due to an increased point defect mobility \cite{2015-Zhang, 2016-Ullah}, which means that fewer defects are created per recoil event.
During the first cascades, mainly point defects occur in both structures, while the concentration of dislocation-network-like structures is still small.

In the second stage, with an increase of irradiation damage, the defect concentration in pure Ni increases and reaches the same magnitude as in the high-entropy structure.
It can be seen that apart from vacancies and stacking-fault tetrahedra, a rather large and stable dislocation loop is formed by the agglomeration of small defects during successive cascades.
In general, the rate of defect annihilation depends on both the size and the spatial distribution of the defects. Small defects can be annihilated if they are hit by a single recoil event, while larger defect structures are not engulfed completely %
and are therefore more stable. Furthermore, the agglomeration of defects in Ni leads to their localization, while the small defects in the HEA are more homogeneously distributed. This means that the probability of a cascade randomly hitting a defect is higher in the HEA and a steady state of annihilation and creation is reached (see video \verb|cascade-CuNiCoFe.avi| in the \hyperref[supplemental]{supplementary material}). A small number of stacking-fault tetrahedra appear, but seem to be unstable. Additionally, we can observe the formation of complex networks of partial dislocations. Because of the large amount of dislocation junctions, these networks must be sessile. Thus, large movements of these networks must again be a series of annihilation and creation events, and they are therefore also unstable against irradiation.
In Ni, the defects are assimilated into larger structures, such as sessile Frank dislocation loops, before they can be destroyed by the next cascade (see the video \verb|cascade-Ni.avi| in the \hyperref[supplemental]{supplementary material}, starting at 0:40; defect close to the center of the box). As such, Ni possesses a higher volume fraction of undisturbed lattice in which the net defect creation rate is always positive. Additionally, in contrast to the HEA, the stacking-fault tetrahedra are numerous.

In the final stage, we also observe a saturation of the defect concentration in pure Ni, while maintaining the steady state in the alloy system. The development
of a steady state in the elemental structure is contrary to earlier publications \cite{2016b-Granberg}, where a
continuous growth of defect networks has been detected. The reason for a reduced growth rate of defect concentrations in Ni is a simulation size effect \cite{Lev16}, since the earlier simulation used larger simulation cells compared to the current work.

In order to examine the stability of the lattice defects in the absence of irradiation, an MD simulation at \unit[800]{K} for \unit[1]{ns} was performed for the initially segregated HEA. The results reveal neither large movement nor creation or dissolution of a significant number of defects, which indicates that the defects are stable and sessile even at elevated temperatures (see video \verb|reanneal-irradiated-CuNiCoFe.avi| in the \hyperref[supplemental]{supplementary material}).

\subsubsection{Chemical order}

In the following
we analyze whether
high energy recoils are able to initiate local segregation processes,
or if the SRO remains more or less constant after every
subsequent cascade.
Again, we
use the Warren-Cowley parameters $\alpha_{1,2}^{ij}$  for the first and second neighbor
shell to quantify the extent of atomic clustering. Figure~\ref{fig:irrad} shows $\alpha^\text{Cu,Cu}$ for two cells: the HEA with initially random atom distribution, and the HEA after MC/MD at \unit[400]{K} which shows segregation tendencies for Cu. The analysis is
limited to atoms in perfect FCC environments and therefore excludes
atoms in defective sites introduced by the irradiation. Order parameters for pairs other than Cu--Cu are plotted in Fig.~S1 in the \hyperref[supplemental]{supplementary material}.
We find that the order parameter $\alpha_1^\text{Cu,Cu}$, although exhibiting initial partial segregation in the structure pre-ordered by MC/MD, eventually converges to small negative values. The initially random structure converges to the same value. Mid-range order is always disturbed by irradiation: the parameter $\alpha_2^\text{Cu,Cu}$ converges to zero in both cases.
This shows that the cascade has two counteracting effects: Firstly, it locally activates the thermodynamically preferrable segregation of Cu. Secondly, it concurrently randomizes the element distribution. The resulting steady state is reached after approximately $700$ cascades.

Figure~\ref{fig:irrad-final-state} shows the comparison of the Warren-Cowley parameters $\alpha_{1,2}$ for all pairs in the CuNiCoFe alloy for both structures (pre-ordered and fully random) after the full irradiation dose. Comparing these results to those in Fig.~\ref{fig:WC_VCSGC} (note the scale difference in the $y$ axes), we observe that the absolute values of
both parameters $\alpha_{1}$ and $\alpha_{2}$ are reduced by roughly one order of magnitude due to the irradiation. Importantly, after irradiation all pairs have practically the same  $\alpha$ parameters regardless of the initial state.  We also note that the $\alpha_2$ values are consistently smaller than the $\alpha_1$, indicating that the ordering effects
under irradiation are limited to the nearest-neighbor shell.

Based on this analysis, we conclude that while collisional cascades provide activation energy for demixing, the system is dynamically driven to a mostly random steady state, independent of the initial chemical order.

\section{Post-Irradiation effects}

\begin{table}
  \centering
  \caption{Composition inside defects of the irradiated sample before
    and after an additional, subsequent MC/MD run. Irradiated sample
    averaged over 50 snapshots (500 cascades). Note that the total
    composition of the sample stays equimolar during MC/MD because of
    the variance constraint and that the composition on the intact
    lattice sites also stays roughly equimolar because of the
    comparatively low amount of defect sites.}
  \label{tab:defect-composition-VCSGC}
  \begin{ruledtabular}
  \begin{tabular}{c
                  >{\centering\arraybackslash}m{0.4\linewidth}
                  >{\centering\arraybackslash}m{0.4\linewidth}}
    element & composition after irradiation
            & composition after subsequent MC/MD \\
    \hline
    \rule{0pt}{2.6ex}%
    Cu & $26.9\%$ & $49.4\%$ \\
    Ni & $24.9\%$ & $17.1\%$ \\
    Co & $23.3\%$ & $14.0\%$ \\
    Fe & $24.9\%$ & $19.5\%$ \\
  \end{tabular}
  \end{ruledtabular}
\end{table}

\begin{figure}
  \centering
  \makebox[\linewidth]{%
    \includegraphics[]{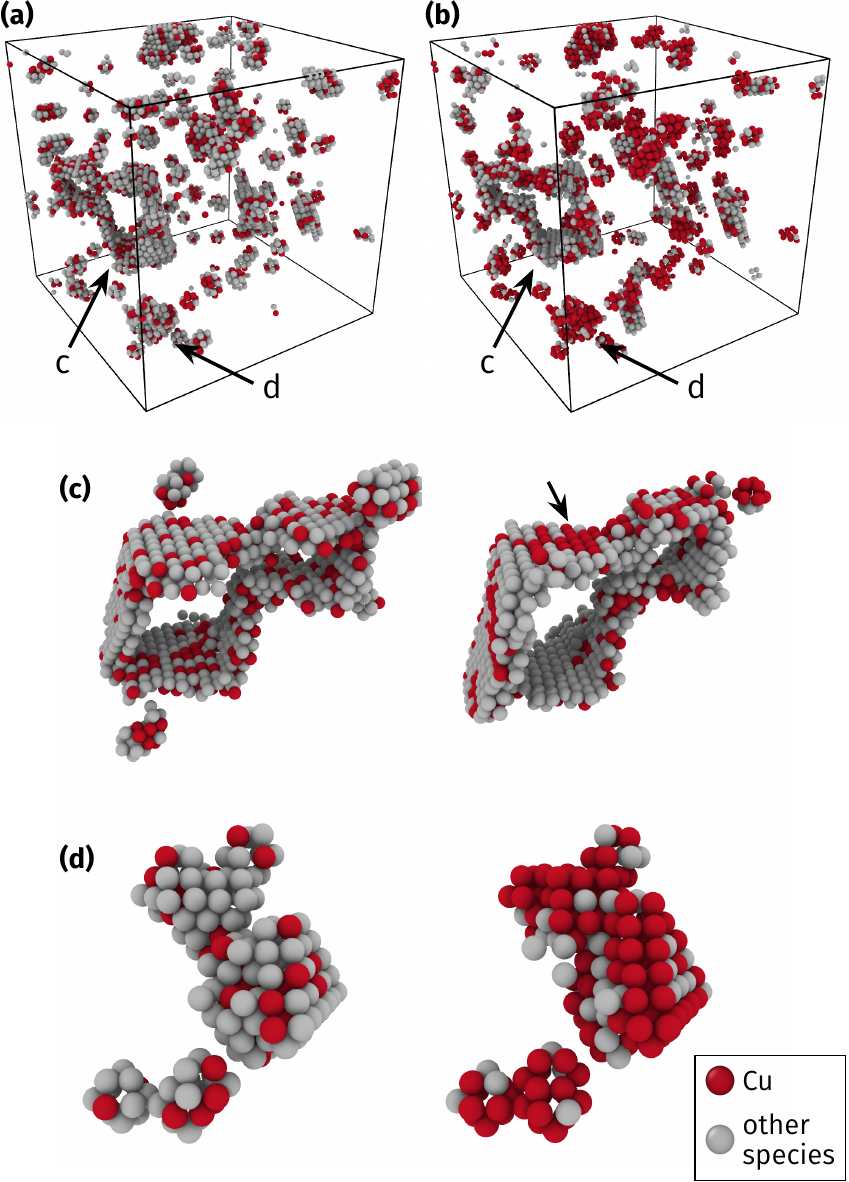}}
  \caption{VC-SGC simulations of the HEA sample after irradiation. (a) Atoms identified as defective by CNA after irradiation [same structure as the righmost snaphost in Fig.~\ref{fig:irrad-dxa}(a)]. Copper atoms are shown in red, all others in gray. (b) The same after subsequent simulation in the VC-SGC ensemble. The concentration of copper atoms around the defects is highly increased. Detailed views of a dislocation structure (c) and a stacking-fault tetrahedron (d) are provided.}
  \label{fig:snapshots-defect-composition}
\end{figure}

Thermodynamic equilibration of a four-component \mbox{CuNiCoFe} alloy with a perfect FCC lattice showed a segregation tendency for Cu atoms. From the results of Sec.~\ref{sec:res:equil} we can assume
that there is not only a chemical
but also a sterical driving force for Cu to segregate from the solid solution into small clusters. We already concluded that the
growth of Cu clusters is accompanied by an increasing Cu atomic volume. Nevertheless, thus far, we mainly concentrated on the thermodynamics of the defect-free system.
Therefore, we again ran hybrid MC/MD simulations in the VC-SGC ensemble.
However, this time, the irradiated sample with pre-existing defects was taken as a starting configuration.
Table~\ref{tab:defect-composition-VCSGC} lists the composition inside the lattice defects. Obviously, defect sites that offer excess volume serve as sinks for Cu atoms and even enhance
the amount of segregated Cu after equilibration. In this regard, the presence of defects enhances the driving
force for local decomposition. Within the defects, we can find a Cu proportion which is approximately \unit[50]{$\%$}. In contrast, during the irradiation the distribution of Cu atoms does not alter significantly.
In Fig.~\ref{fig:snapshots-defect-composition} we visualized the atoms located within the defect structure before (left) and after (right) the equilibration. Copper atoms are displayed with a red color, while all others are shown in gray. Figures \ref{fig:snapshots-defect-composition}c and d highlight two defects separately. From these illustrations we draw
the conclusion that Cu preferentially segregates on, e.g., the edges of stacking-fault tetrahedra rather than in the stacking faults themselves. The reason lies in the higher local excess volume of these sites (cf.\ Fig.~\ref{fig:ScatterPlotVoronoiVolType}).

In conclusion, all results suggest that the mixing enthalpy of Cu is the decisive factor for phase stability in this case, exceeding the total entropy with decreasing temperature.  The limited segregation tendencies of the perfect crystal are surpassed in the presence of defects, which act as sinks for copper.

\section{Conclusion}
\label{sec4}

Using a model CuNiCoFe HEA with a tendency for copper segregation, we observe that the system reaches a steady state of defect concentration and of chemical order under irradiation, irrespective of the initial structure. In contrast to pure metals, irradiation leads to less mobile point defects and therefore a larger number of isolated defects instead of recombination or agglomeration. While, e.g., Ni increases its defect concentration continuously with increasing irradiation dose due to the agglomeration of defects in larger, more stable structures, the HEA quickly reaches a steady state of defect creation and annihilation. Irradiation provides the thermal activation for demixing of copper, but at the same time re-randomizes the elemental distribution. The resulting steady state of chemical order is close to a random solid solution, but still shows traces of local precipitation. Furthermore, our simulations reveal that various lattice defects act as sinks for copper, most likely since the copper atoms are the largest species and the defects provide excess volume. This clustering of copper at defects is also suppressed during irradiation for the reasons enumerated above. Together, these effects explain the high irradiation resistance of HEAs.

\section*{Supplementary material}
\label{supplemental}

See \href{https://doi.org/10.1063/1.4990950}{supplementary material} for videos of the irradiation simulations,
a figure of the detailed evolution of SRO during irradiation, and
additional data for the verification of the EAM potential.

\begin{acknowledgments}
The authors gratefully acknowledge travel grants through the PPP Finland program of the Deut\-scher Aka\-de\-mi\-scher Aus\-tausch\-dienst (DAAD) and K.A. acknowledges financial support by the Deutsche Forschungsgemeinschaft (DFG) through project Grant No.\ STU 611/2-1. Computing time was made available by the Technische Universit\"at Darmstadt on the Lichtenberg cluster and the IT Center for Science, CSC, Finland. \end{acknowledgments}

\appendix

\begin{table*}
  \centering
  \caption{Comparison between the potential and literature values for
    properties of Cu, Ni, Co, and Fe. The table shows lattice
    constants $a$ and $c$, the cohesive energy $E_\text{coh}$, the
    components of the stiffness tensor $c_{ij}$, the stacking fault
    energy $\gamma_\text{SF}$, the unstable stacking fault energy
    $\gamma_\text{USF}$, the critical shear stress $\sigma_\text{SF}$
    for a stacking fault, the twinning fault energy
    $\gamma_\text{TF}$, the unstable twinning fault energy
    $\gamma_\text{UTF}$, and the melting point $T_\text{melt}$.}
  \label{tab:validation-elements}
  \begin{ruledtabular}
  \begin{tabular}{lcc@{\hskip 0em}lcc@{\hskip 0em}lcc@{\hskip 0em}lcc@{\hskip 0em}lcc@{\hskip 0em}lcc@{\hskip 0em}l}
    & \multicolumn{3}{c}{Cu (FCC)}
    & \multicolumn{3}{c}{Ni (FCC)}
    & \multicolumn{3}{c}{Co (FCC)}
    & \multicolumn{3}{c}{Co (HCP)}
    & \multicolumn{3}{c}{Fe (FCC)}
    & \multicolumn{3}{c}{Fe (BCC)} \\
    \cline{2-4}
    \cline{5-7}
    \cline{8-10}
    \cline{11-13}
    \cline{14-16}
    \cline{17-19}
    \\[-8pt] %
    & Pot. & \multicolumn{2}{c}{Ref.}
    & Pot. & \multicolumn{2}{c}{Ref.}
    & Pot. & \multicolumn{2}{c}{Ref.}
    & Pot. & \multicolumn{2}{c}{Ref.}
    & Pot. & \multicolumn{2}{c}{Ref.}
    & Pot. & \multicolumn{2}{c}{Ref.} \\
    \hline
    \rule{0pt}{2.6ex}%
    $a$ (\AA) & 3.615 & 3.615 & \cite{Lide}
              & 3.520 & 3.524 & \cite{Lide}
              & 3.549 & 3.545 & \cite{Lide}
              & 2.501 & 2.507 & \cite{Lide}
              & 3.628 & 3.647 & \cite{Lide}
              & 2.866 & 2.867 & \cite{Lide} \\
    $c$ (\AA) &       & &
              &       & &
              &       & &
              & 4.076 & 4.069 & \cite{Lide}
              &       & &
              &       & & \\
    $c/a$ &       & &
          &       & &
          &       & &
          & 1.630 & 1.623 & \cite{Lide}
          &       & &
          &       & & \\
    $E_\text{coh}$ (eV/at.) & $-3.54$ & $-3.49$ & \cite{Kittel}
                           & $-4.45$ & $-4.44$ & \cite{Kittel}
                           & $-4.40$ & $-4.29$ &
                                       \footnote{There are
                                       considerably different absolute values
                                       reported in literature. We
                                       assumed a difference of
                                       $0.1\,$eV/atom to the HCP
                                       structure, while literature
                                       reports differences from
                                       $0.02\,$eV/atom to
                                       $0.15\,$eV/atom.}%
                                       $^,$%
                                       \cite{Cardellini1993, PenaOShea2010}
                           & $-4.41$ & $-4.39$ & \cite{Kittel}
                           & $-4.20$ & $-4.17$ & \cite{Mueller2007}
                           & $-4.29$ & $-4.28$ & \cite{Kittel} \\
    $c_{11}$ (GPa) & 170 & 168 & \cite{Lide}
                  & 246 & 248 & \cite{Lide}
                  & 213 & 225 & \cite{Gump1999}
                  & 263 & 307 & \cite{Lide}
                  & 107 & 154 & \cite{Zarestky1987}
                  & 229 & 226 & \cite{Lide} \\
    $c_{12}$ (GPa) & 122 & 122 & \cite{Lide}
                  & 147 & 155 & \cite{Lide}
                  & 157 & 160 & \cite{Gump1999}
                  & 158 & 165 & \cite{Lide}
                  & \phantom{0}98 & 122 & \cite{Zarestky1987}
                  & 136 & 140 & \cite{Lide} \\
    $c_{13}$ (GPa) & & &
                  & & &
                  & & &
                  & 124 & 103 & \cite{Lide}
                  & & &
                  & & & \\
    $c_{33}$ (GPa) & & &
                  & & &
                  & & &
                  & 363 & 358 & \cite{Lide}
                  & & &
                  & & & \\
    $c_{44}$ (GPa) & \phantom{0}76 & \phantom{0}76 & \cite{Lide}
                  & 125 & 124 & \cite{Lide}
                  & \phantom{0}99 & \phantom{0}92 & \cite{Gump1999}
                  & \phantom{0}65 & \phantom{0}76 & \cite{Lide}
                  & \phantom{0}80 & \phantom{0}77 & \cite{Zarestky1987}
                  & 117 & 116 & \cite{Lide} \\
    $\gamma_\text{SF}$ (mJ/m$^2$) & \phantom{0}22 & 36--49
                                                 & \cite{Ogata2002, Jin2008, Zimmerman2000}
                                & \phantom{0}97 & 133--183
                                                & \cite{Jin2008, Zimmerman2000}
                                & $-40$ & &
                                & & &
                                & & &
                                & & & \\
    $\gamma_\text{USF}$ (mJ/m$^2$) & 110 & 95--210 &
                                          \footnote{Value for Ref.~\onlinecite{Ogata2002}
                                           computed by
                                           numeric integration of the
                                           stress data.}%
                                       $^,$%
                                       \cite{Ogata2002,Jin2008,Zimmerman2000}
                                 & 251 & 258
                                       & \cite{Jin2008}
                                 & 174 & &
                                 & & &
                                 & & &
                                 & & & \\
    $\sigma_\text{SF}$ (GPa) & 2.3 & 2.2 & \cite{Ogata2002}
                            & 5.1 & &
                            & 3.7 & &
                            & & &
                            & & &
                            & & & \\
    $\gamma_\text{TF}$ (mJ/m$^2$) & \phantom{0}23 & &
                                & \phantom{0}99 & &
                                & $-44$ & &
                                & & &
                                & & &
                                & & & \\
    $\gamma_\text{UTF}$ (mJ/m$^2$) & 121 & 143 & \cite{Jin2008}
                                 & 298 & 186 & \cite{Jin2008}
                                 & 159 & &
                                 & & &
                                 & & &
                                 & & & \\
    $T_\text{melt}$ (K) & 1175 & 1357 & \cite{Chakrabarti1991a, Chakrabarti1994a}
                       & 1513 & 1728 & \cite{Chakrabarti1991a, Chakrabarti1994a}
                       & & &
                       & 1756 & 1768 & \cite{Nishizawa1984a}
                       & & &
                       & 2044 & 1811 & \cite{Swartzendruber1991a} \\
  \end{tabular}
  \end{ruledtabular}
\end{table*}

\section{Validation of the EAM potential}
\label{sec:pot-val}

We used an EAM potential by Zhou \textit{et al.} \cite{Misfit} for the Cu-Ni-Co-Fe system. This potential was tested for HEAs under driven conditions: Thin-film growth by sputtering yields comparable structures to experiment and the BCC to FCC transition is reproduced correctly \cite{Xie2013,XieBraTho14}. Furthermore, damage accumulation during irradiation agrees with data obtained by Rutherford backscattering spectrometry \cite{Zha16c}.

We performed additional validation of the potential by comparing the properties of the elements and of binary mixtures to reference values. In a first step, we produced FCC lattices of Cu, Ni, Co, and Fe, as well as HCP Co and BCC Fe. These were minimized to obtain the \unit[0]{K} ground state values of the lattice constants $a$ and $c$ and the cohesive energy $E_\text{coh}$. The comparison to literature values in Table~\ref{tab:validation-elements} reveals a good match. We then obtained the stiffness tensor using Hooke's law by calculating the stress tensors for finite deformations of the box at 1\% strain. Apart from HCP cobalt, which does not occur in our samples, and FCC iron, which is too soft, the reference values are reproduced quite well. Using the procedure described in Ref.~\onlinecite{Ogata2002}, we calculated the stacking fault energy $\gamma_\text{SF}$, the unstable stacking fault energy $\gamma_\text{USF}$, and the critical shear stress $\sigma_\text{SF}$ for a stacking fault. While literature data contains large uncertainties and variations, the general trend of available data agrees with the predictions of the potential. As far as reference data is available, the same is true for the twinning faults, which we calculated using the method described in Ref.~\onlinecite{Tadmor2003}. The negative values for the stacking and twinning faults in Co are of course the result of the lower energy of the stable HCP structure.

Furthermore, we calculated the elastic constants of those binary alloys, for which reference data is available, i.e., in the Fe--Co--Ni system. The results are presented in Table~S.1 in the \hyperref[supplemental]{supplementary material}. Since the cross terms of the potential are obtained through a mixing procedure, the results match the expectations from the properties of the elements: The elastic constants match well, except for alloys containing FCC iron or HCP cobalt, which are too soft.  To additionally determine the melting points of the elements and some miscible alloys, liquid--solid interface simulations were carried out \cite{Ercolessi1993}. A simulation cell of 2000 atoms was relaxed with the correct crystal structure and lattice constant, without pressure control. An equally sized box was molten and cooled down to the desired temperature, with pressure control in the $z$ direction. The two boxes where then combined in the $z$ direction, and were let to relax with pressure control in all directions. Below the melting point, the cell will solidify and the volume of the cell will decrease, while the opposite will happen above the melting point. Due to the chosen temperature stepping, an error of around \unit[25]{K} is expected. The results are presented in Table~S.2 in the \hyperref[supplemental]{supplementary material} and show that the melting points slightly deviate from the reference values, but capture the features of the phase diagrams: For the pure elements the deviation of the melting point is below 15\% of the absolute value. The melting point of HCP cobalt agrees with the literature value within the expected error of the simulation. The melting points of the different binary subsystems studied are all within 10\% of the expected value.  We point out that a good description of the melting points is particularly important for the irradiation simulations, since several previous studies show that the outcome of molecular dynamics of collision cascades depends directly on the melting point \cite{Nor98,Nor01}.

In a next step, random solid solutions of all binary subsystems on FCC lattices were prepared. Additionally, if the binary alloy contained Fe or Co, BCC or HCP lattices, respectivley, were created. Different molar fractions of each alloy and lattice type were minimized to obtain the mixing enthalpy
\begin{equation}
\label{eq:mixing_enthalpy}
H_\mathrm{M} = H_\mathrm{AB} - (x_\mathrm{A} H_\mathrm{A} + x_\mathrm{B} H_\mathrm{B}),
\end{equation}
where \(x_\mathrm{A}\), \(x_\mathrm{B}\), \(H_\mathrm{A}\), and \(H_\mathrm{B}\) are the molar fractions and enthalpies of the constituents A and B, and where $H_\mathrm{AB}$ is the total enthalpy of the alloy. We
included mixing enthalpy curves of all binary subsystems with respect to the concentration into Figs.~S.2--S.7 in the \hyperref[supplemental]{supplementary material} and compared them to
literature values. We additionally compared our results to data from the \textsc{thermo-calc} software and database \cite{ThermoCalc2017a, Andersson2002}. In most cases the EAM potential correctly describes the trend observed
in literature. However, it should be noted that even different literature values do not completely agree and that there is a large spread for some systems.

\begin{table}
  \centering
  \caption{Lattice constants of binary random solid solutions at \unit[300]{K}. The lattice structure is FCC unless stated otherwise.}
  \label{tab:binary_lattice_constants}
  \begin{ruledtabular}
  \begin{tabular}{l c r@{\hskip 0em} l}
     & \multicolumn{3}{c}{\(a\) (\AA)} \\
        \cline{2-4}
        \\[-8pt]
     & Pot. & \multicolumn{2}{c}{Ref.} \\

    \hline
    \rule{0pt}{2.6ex}%
        \(\mathrm{Cu_{3}Ni_{97}}\) & 3.536 & 3.526 & \cite{Coles1956a,Clarke1971a} \\
        \(\mathrm{Cu_{49}Ni_{51}}\) & 3.591 & 3.564 & \cite{Coles1956a,Clarke1971a} \\
        \(\mathrm{Cu_{95}Ni_{5}}\) & 3.643 & 3.609 & \cite{Coles1956a,Clarke1971a}	\\

        \cline{1-4}
\rule{0pt}{2.6ex}%
        \(\mathrm{Cu_{8}Co_{92}}\) & 3.568 & 3.551 & \cite{Predel1993a} 	\\
        \(\mathrm{Cu_{21}Co_{79}}\) & 3.607 & 3.577 & \cite{Predel1993a} 	\\
        \(\mathrm{Cu_{86}Co_{14}}\) & 3.636 & 3.668 & \cite{Predel1993a} 	\\
        \(\mathrm{Cu_{99}Co_{1}}\) & 3.646 & 3.674 & \cite{Predel1993a} 	\\

        \cline{1-4}
\rule{0pt}{2.6ex}%
	\(\mathrm{Cu_{23}Fe_{77}}\) (BCC)	& 2.890 & 2.890 & \cite{Klement1965a} 	\\
        \(\mathrm{Cu_{96}Fe_{4}}\) & 3.648 & 3.616 & \cite{Klement1965a}	\\

        \cline{1-4}
\rule{0pt}{2.6ex}%
        \(\mathrm{Ni_{26}Co_{74}}\) & 3.555 & 3.539 & \cite{Nishizawa1983a} \\
        \(\mathrm{Ni_{29}Co_{71}}\) & 3.554 & 3.538 & \cite{Nishizawa1983a} \\
        \(\mathrm{Ni_{49}Co_{51}}\) & 3.548 & 3.533 & \cite{Nishizawa1983a} \\
        \(\mathrm{Ni_{51}Co_{49}}\) & 3.548 & 3.533 & \cite{Nishizawa1983a} \\
        \(\mathrm{Ni_{72}Co_{28}}\) & 3.541 & 3.528 & \cite{Nishizawa1983a} \\

        \cline{1-4}
\rule{0pt}{2.6ex}%
        \(\mathrm{Ni_{18}Fe_{82}}\) (BCC)	& 2.869 & 2.867 & \cite{Swartzendruber1991a} \\
        \(\mathrm{Ni_{39}Fe_{61}}\) & 3.594 & 3.595 & \cite{Swartzendruber1991a} \\
        \(\mathrm{Ni_{48}Fe_{52}}\) & 3.584 & 3.590 & \cite{Swartzendruber1991a} \\
        \(\mathrm{Ni_{49}Fe_{51}}\) & 3.582 & 3.587 & \cite{Swartzendruber1991a} \\
        \(\mathrm{Ni_{78}Fe_{22}}\) & 3.552 & 3.552 & \cite{Swartzendruber1991a} \\

        \cline{1-4}
\rule{0pt}{2.6ex}%
   	\(\mathrm{Co_{11}Fe_{89}}\) (BCC)	& 2.869 & 2.861 & \cite{Ellis1941a} 	\\
    	\(\mathrm{Co_{24}Fe_{76}}\) & 2.857 & 2.860 & \cite{Ellis1941a} 	\\
  \end{tabular}
  \end{ruledtabular}
\end{table}

In order to get a clearer picture, we investigated the miscibility gaps of the binary subsystems. For this, equimolar alloys on FCC lattices were equilibrated in the semi-grand-canonical ensemble without variance constraint. By systematically varying $\Delta\mu$, final compositions ranging over the whole phase diagram were obtained. The simulations were performed at 300\,K and 800\,K.
The concentration as a function of the chemical potential difference can be used to identify miscibility gaps in a binary subsystem: Given a dense mesh of \(\Delta\mu_\mathrm{A-B}\) values, miscibility gaps appear as discontinuities in the \(x_\mathrm{B}(\Delta\mu_\mathrm{A-B})\) curves. Performing this analysis and comparing it to reference data showed that the potential is mostly able to reproduced phase diagram features, even though the miscibility is generally a bit too high. In the Co-containing systems, this method will not find the HCP phase, since the simulation setup was always initiated from an FCC structure. Due to the difference in unit cell shape, the simulation suppresses the transition to HCP. We omit this investigation, since the HEA remains in the FCC structure in all conditions. The \(x_\mathrm{B}(\Delta\mu_\mathrm{A-B})\) curves and a more detailed discussion of all binary subsystems can be found in Figs.~S.8--S.13 in the \hyperref[supplemental]{supplementary material}.

Finally, the lattice constants of the binary systems were calculated from these simulations by time averaging the simulation cell volume of the different binary systems at 300\,K. The data is shown in Table~\ref{tab:binary_lattice_constants} and is in good agreement with experimental data. This confirms that the cross terms of the potential correctly predict the atomic volumes in alloys.

All in all, the potential performs reasonably well, reproducing the correct stable phases, densities, stiffnesses, and melting points. The limited data for stacking fault and twinning-fault energies suggests that lattice defects should be correctly reproduced. The main weakness of the potential is that the solubility of the alloys is too high in some cases, meaning that the tendencies for copper precipitation may be underestimated in the present work.

\clearpage


\section*{Supplementary material}

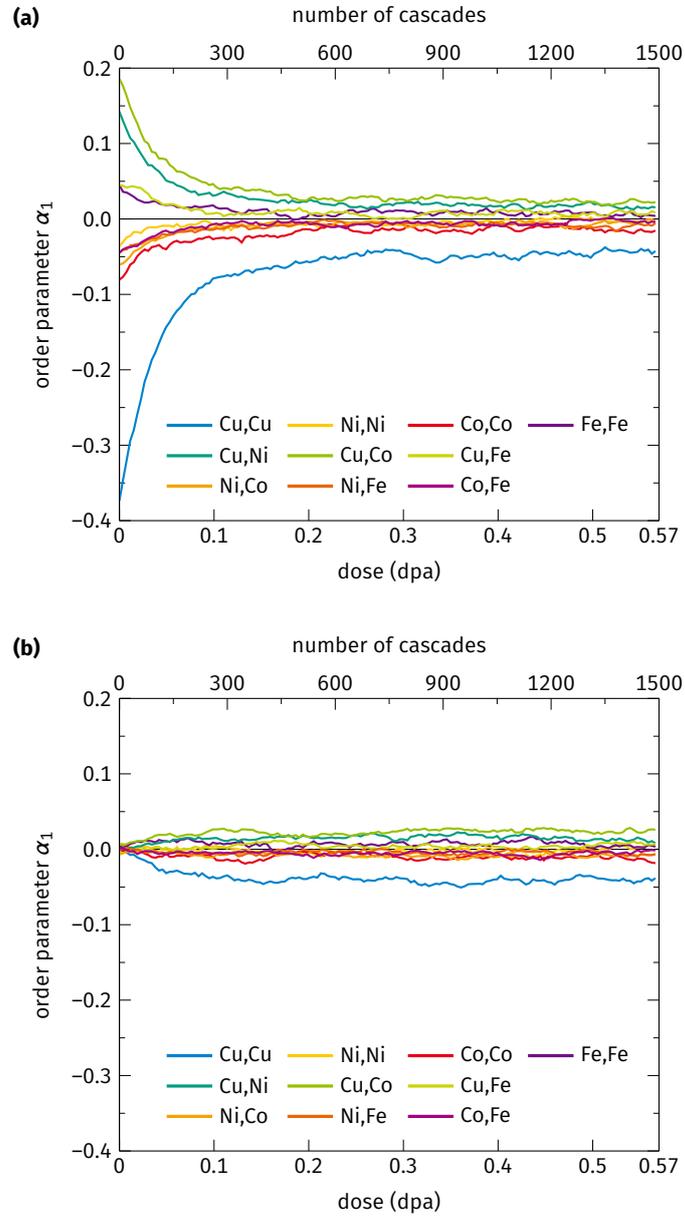

**Figure S.1:** Evolution of the SRO in CuNiCoFe HEAs during irradiation for all pairs (cf. Fig. 5 in the main paper). (a) Starting from the ordered sample (MC/MD at 400 K). (b) Starting from the initially random sample.

1

**Table S1:** Elastic constants of binary alloys. FCC and HCP represent random solid solutions on these lattices. B2 represents an ordered CsCl structure. The stiffness tensor $c_{ij}$, the bulk modulus $K$, the shear modulus $G$, and the Young's modulus $Y$ are listed. For the shear and Young's moduli, we assumed that the reference is for a polycrystalline sample. Thus, we calculated the Voigt–Reuss–Hill (VRH) average from the stiffness tensor.[1–3]

| | $c_{11}$ (GPa) | | $c_{12}$ (GPa) | | $c_{13}$ (GPa) | | $c_{33}$ (GPa) | | $c_{44}$ (GPa) | | $K$ (GPa) | | $G_{VRH}$ (GPa) | | $Y_{VRH}$ (GPa) | |
|---|---|---|---|---|---|---|---|---|---|---|---|---|---|---|---|---|
| | Pot. | Ref. | Pot. | Ref. | Pot. | Ref. | Pot. | Ref. | Pot. | Ref. | Pot. | Ref. | Pot. | Ref. | Pot. | Ref. |
| Co50Ni50 (FCC) | 231 | | 153 | | | | | | 113 | | 179 | 257[4,a] | 74 | | 194 | |
| Co68Ni32 (FCC) | 225 | 239[5] | 155 | 155[5] | | | | | 108 | 132[5] | | | | | | |
| Co68Ni32 (HCP) | 278 | 326[5] | 153 | 161[5] | 108 | | 323 | 358[5] | 61 | 74[5] | | | | | | |
| Fe30Ni70 (FCC) | 194 | 233[6] | 138 | 146[6] | | | | | 117 | 127[6] | 156 | | 67 | 78[6] | 175 | 206[6] |
| Fe50Ni50 (FCC) | 164 | 200[6] | 130 | 142[6] | | | | | 108 | 107[6] | 142 | | 53 | 64[6] | 141 | 171[6] |
| Fe65Ni35 (FCC) | 149 | 215[7] | 129 | 175[7] | | | | | 100 | | 136 | 188[7] | 43 | 57–59[6] | 117 | 145–151[6] |
| Fe70Ni30 (FCC) | 144 | 146[6] | 127 | 88[6] | | | | | 98 | 113[6] | 133 | 110;180[8,b] | 40 | 65[6] | 110 | 161[6] |
| FeCo (B2) | 222 | | 171 | | | | | | 129 | | 188 | 247[9,c] | 69 | | 183 | |

[a] The authors of Ref. 4 (a DFT calculation) note that the stiffness of pure Ni lies around 40% higher than the available experimental data. If we assume that the same is true for the binary alloy, the resulting value of $K = 184$ GPa compares favorably to the potential.

[b] The lower value corresponds to the ferromagnetic, the higher value to the paramagnetic phase.

[c] The DFT calculations in Ref. 9 again overestimate the stiffness of pure Fe by around 40%. Correcting the value for FeCo by the same percentage yields $K = 176$ GPa, which is in good agreement with the potential.

**Table S.2:** Melting points $T_\text{melt}$ of the elements and the miscible binary alloys.

| | $T_\text{melt}$ (K) | |
|---|---|---|
| | Pot. | Ref. |
| Ni | 1513 | 1728 [10,11] |
| Cu | 1175 | 1357 [10,11] |
| Fe | 2044 | 1811 [12] |
| Co | 1756 | 1768 [13] |
| NiCo | 1625 | 1748 [13] |
| NiFe | 1656 | 1718 [12] |
| Ni$_2$Fe | 1594 | 1713 [12] |
| Ni$_3$Fe | 1582 | 1714 [12] |

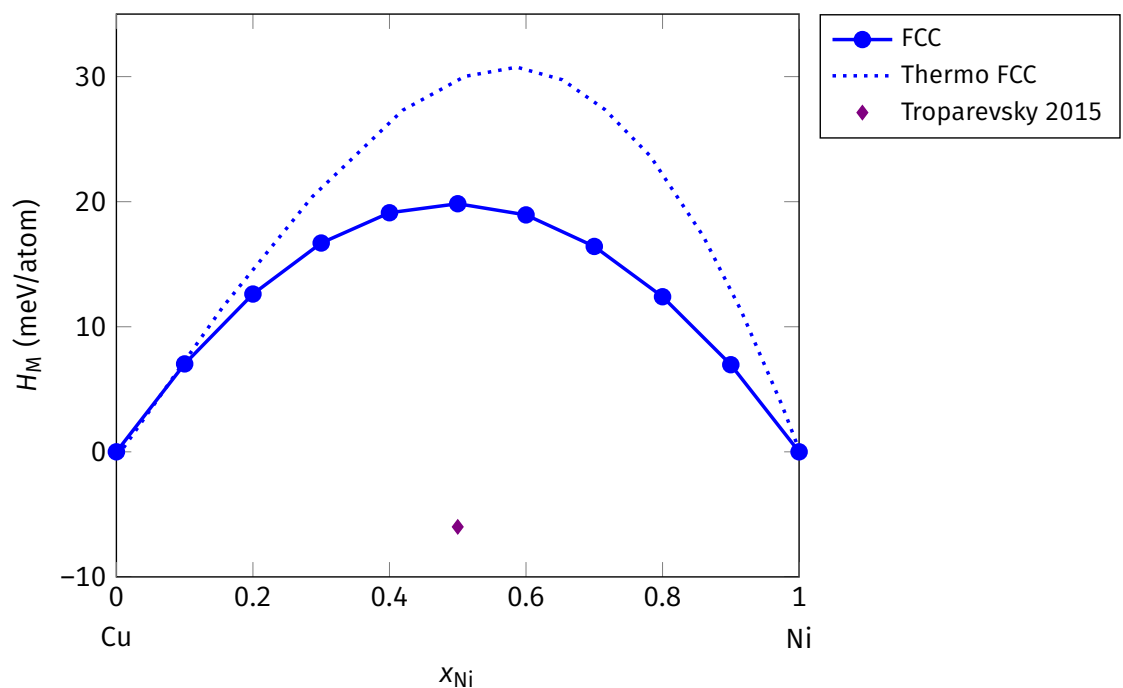

**Figure S.2:** Mixing enthalpy of a Cu–Ni binary subsystem with an FCC (blue dots) lattice structure. We found reference data in Troparevsky 2015[14] (violet diamonds) for an equimolar alloy system. The obtained mixing enthalpy shows a large deviation from our results. Enthalpy data obtained from the THERMO-CALC software[15,16] (dotted blue line), on the other hand, shows a similar trend compared to the present EAM potential.

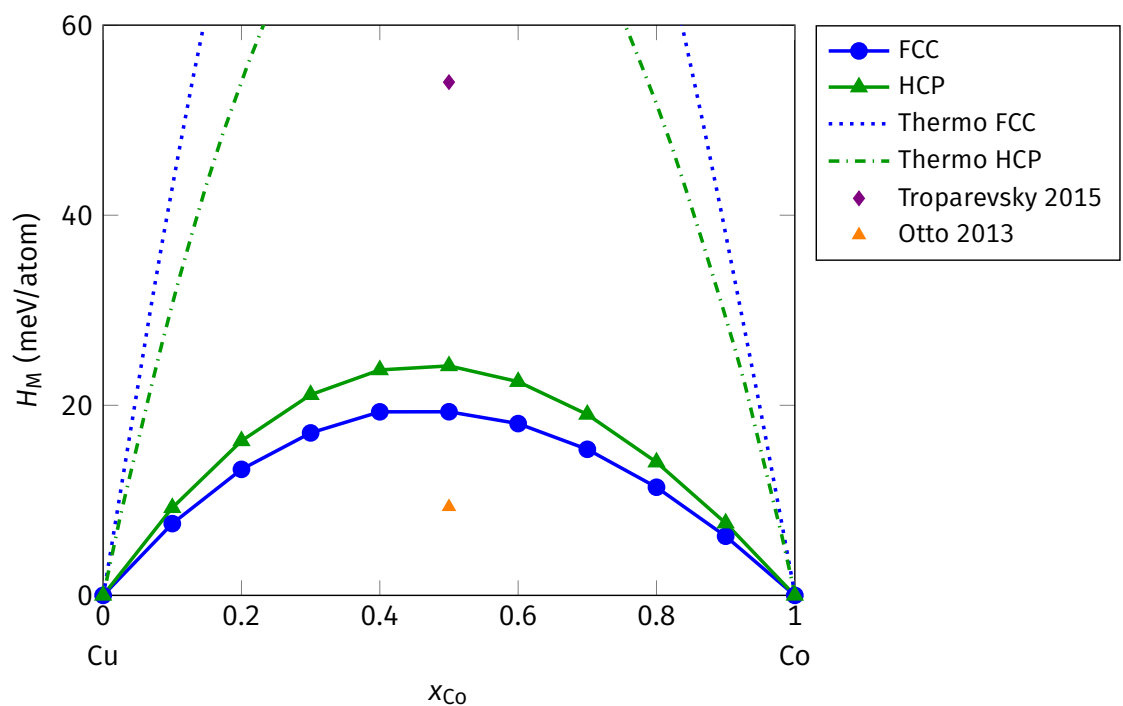

**Figure S.3:** Mixing enthalpies of Cu–Co binary subsystems with FCC (blue dots) and with HCP (green triangles) lattice structures. Reference data from Otto 2013[17] (orange triangles) roughly agrees with the potential, while both Troparevsky 2015[14] (violet diamonds) and THERMO-CALC[15,16] (dotted blue line and dash-dotted green line) specify much higher values.

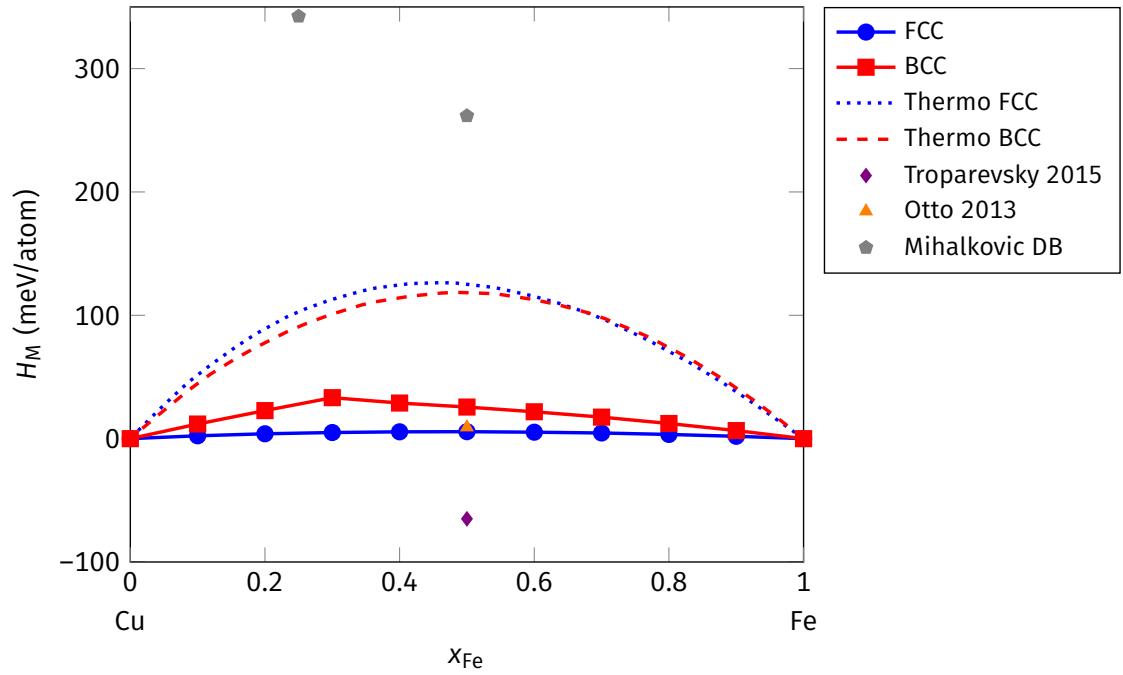

**Figure S.4:** Mixing enthalpies of Cu–Fe binary subsystems with FCC (blue dots) and with BCC (red squares) lattice structures. We compared our mixing enthalpies with Otto 2013[17] (orange triangles), Troparevsky 2015[14] (violet diamonds), the Mihalkovic database[18] (gray pentagons), and the THERMO-CALC database[15,16] (dotted blue line and dashed red line). In this subsystem, too, the value of Otto 2013 fits best with the results of the present study. In this case, the huge spread of the reference data becomes particularly apparent.

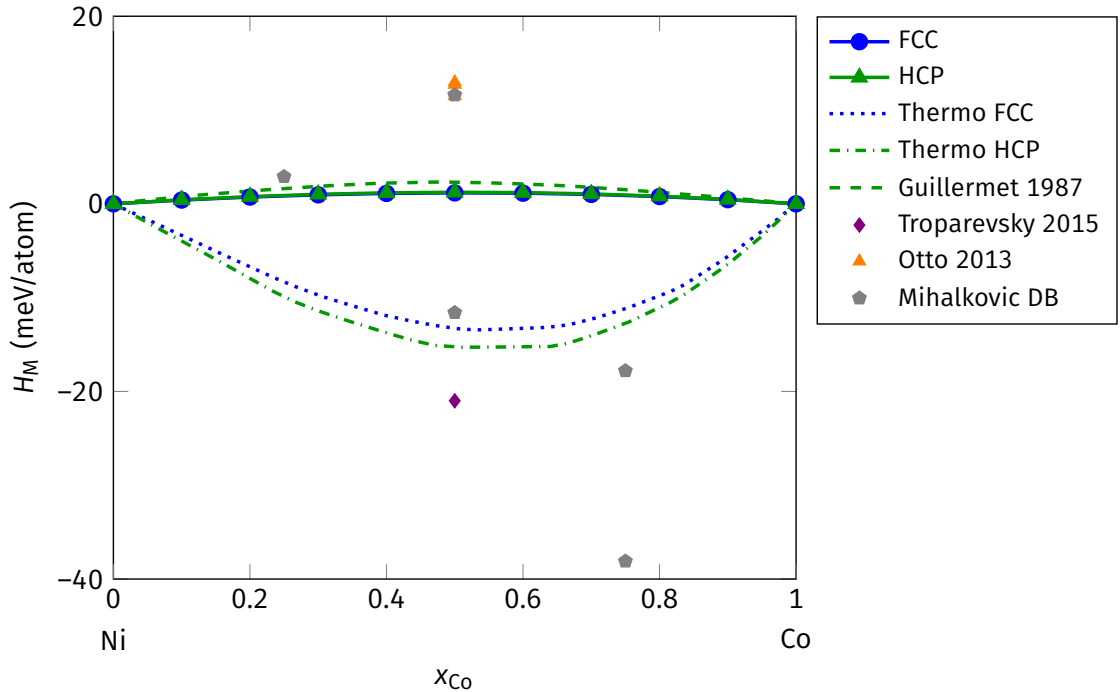

**Figure S.5:** Mixing enthalpies of Ni–Co binary subsystems with FCC (blue dots) and HCP (green triangles) lattice structures, both of which exhibit the same mixing enthalpies. We compared our findings with the data of Otto 2013[17] (orange triangles), Troparevsky 2015[14] (violet diamonds), Guillermet 1987[19] (green dashed line), the Mihalkovic database[18] (gray pentagons), and the THERMO-CALC database[15,16] (dotted blue line and dash-dotted green line). It can be seen that the agreement with the reference values depends on the composition. Here, the values of Guillermet 1987 almost coincide with our data. All other reference data strongly differ, some even by their sign. It should be noted that the positive mixing enthalpies of the EAM potential are so small that the resulting systems are still miscible at room temperature, as expected from experiment.

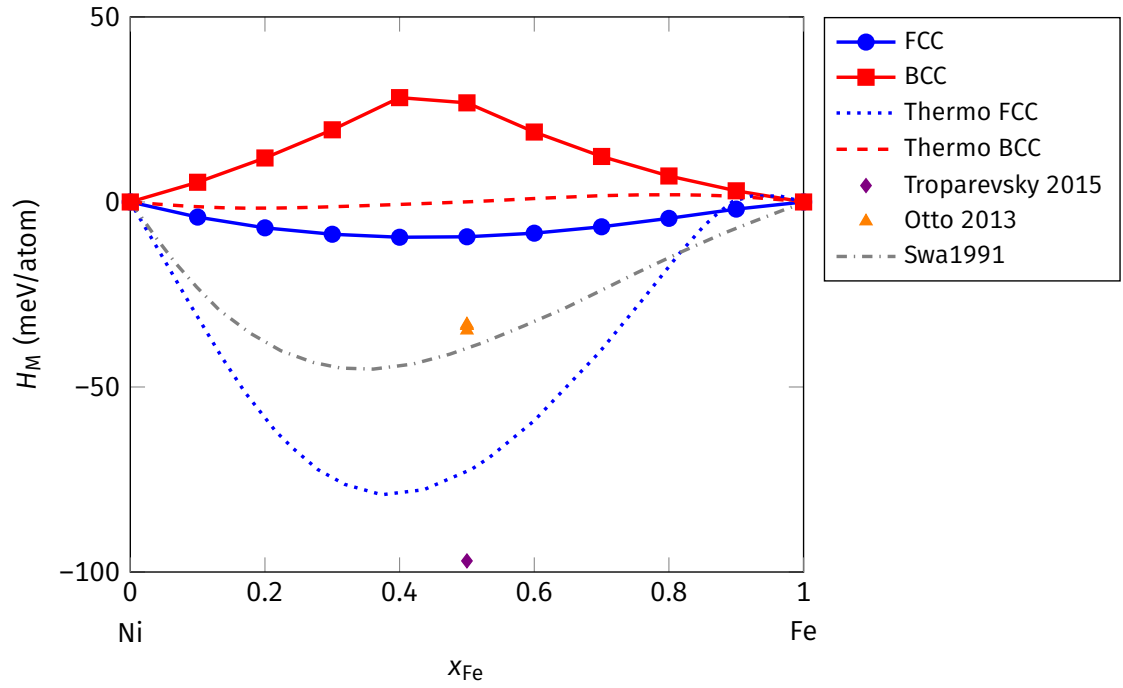

**Figure S.6:** Mixing enthalpies of Ni–Fe subsystems with FCC (blue dots) and BCC (red squares) lattice structures. As a reference, we included data from Swartzendruber 1991[12] (gray dashed line), Otto 2013[17] (orange triangles), Troparevsky 2015[14] (violet diamonds), and the THERMO-CALC database[15,16] (dotted blue line and dashed red line). Again, the absolute values in this subsystem are not comparable. Nevertheless, the right trend is obtained: The system is miscible.

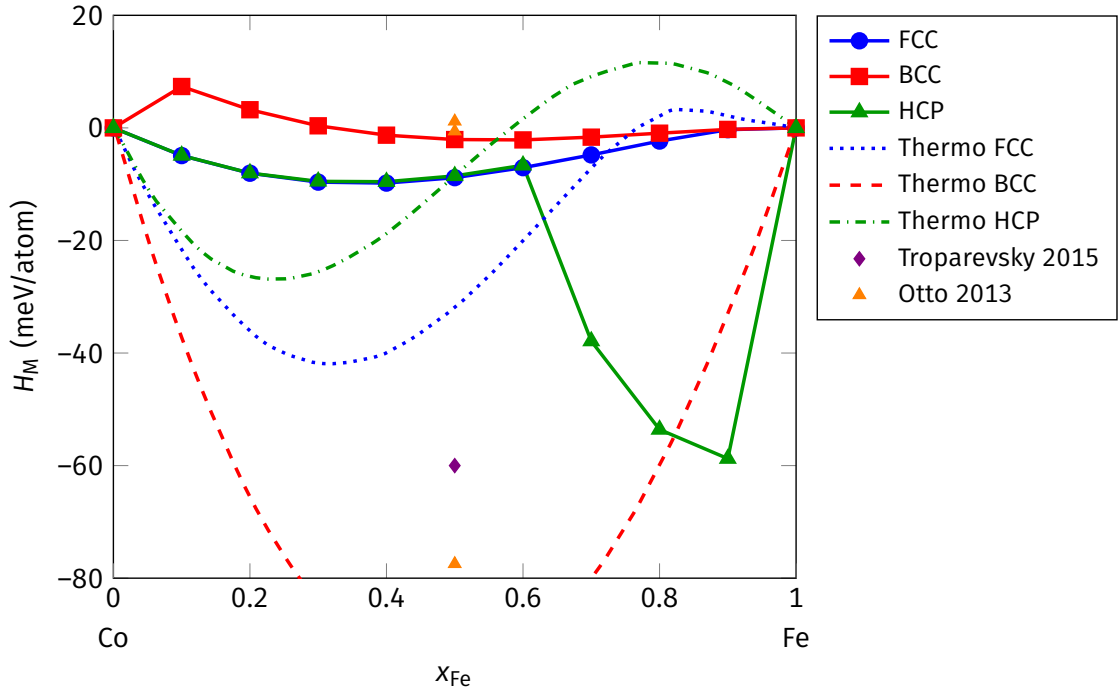

**Figure S.7:** Mixing enthalpy of Co–Fe binary subsystems with FCC (blue dots), HCP (green triangles), and BCC (red squares) lattice structures. We included reference data from Otto 2013[17] (orange triangles) and Troparevsky 2015[14] (violet diamonds) for an equimolar alloy. The data from THERMO-CALC[15,16] (dotted blue line, dashed red line, and dash-dotted green line) deviates strongly from the potential and the other references in this case. Again, the reference data from Otto 2013 almost coincides with our calculations. The system is miscible, as expected from the phase diagram.[20] The anomalous decrease in mixing enthalpies for the simulated, iron-rich HCP structure describes a thermodynamically unstable state. It does not occur in the present work, which is concerned with FCC HEAs.

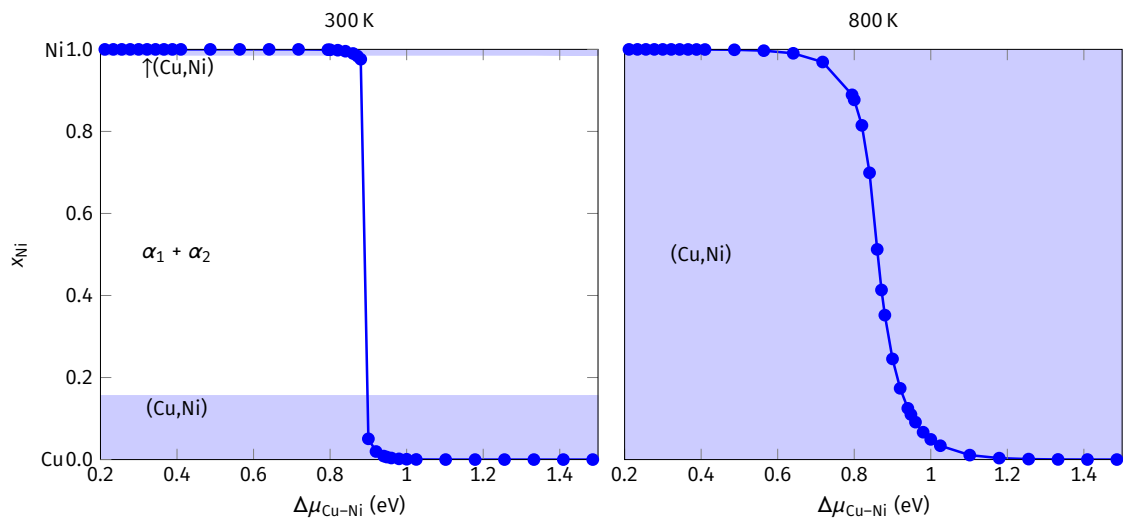

**Figure S.8:** Molar fraction of Ni in the Cu–Ni binary subsystem as a function of the chemical potential difference $\Delta\mu$ at 300 K and 800 K. Shaded areas indicate phases extracted and extrapolated from Chakrabarti *et al.*,[10,11] while white areas are miscibility gaps. At 300 K the miscibility is slightly too low, leading to an overly large miscibility gap (indicated by an absence of data points). An FCC solid solution is observed over the whole compositional range at 800 K, in accordance with the reference phase diagram.

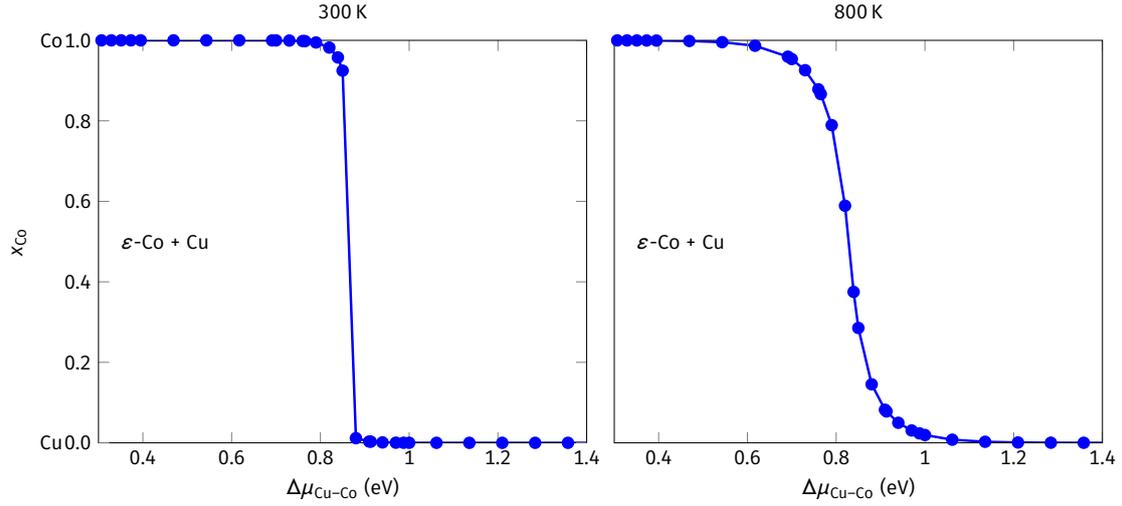

**Figure S.9:** Molar fraction of Co in the Cu–Co binary subsystem as a function of the chemical potential difference $\Delta\mu$ at 300 K and 800 K. From Nishizawa *et al.*[13] it can be seen that Cu–Co only shows insignificant miscibility at 300 K and 800 K. While this is almost reproduced at 300 K with only a slight miscibility of Cu in Co, at 800 K, miscibility is observed over the whole compositional range. Due to the simulation setup, transitions from FCC to HCP cannot be observed, which leads to an FCC solid solution at all concentrations. Possibly, a phase separation into Cu (FCC) and $\varepsilon$-Co (HCP) may be more stable than the FCC solid solution and recover the phase diagram qualitatively. In HEAs, though, cobalt is not known to precipitate in an HCP phase.

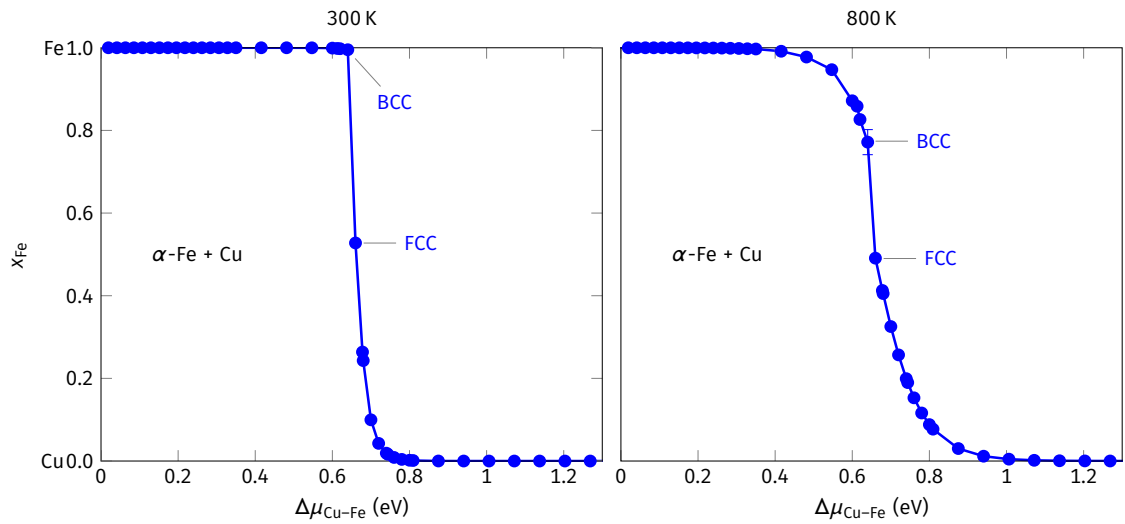

**Figure S.10:** Molar fraction of Cu in the Cu–Fe binary subsystem as a function of the chemical potential difference $\Delta\mu$ at 300 K and 800 K. From Ansara *et al.*[21] it can be seen that Cu–Fe only shows insignificant miscibility at 300 K and 800 K. An FCC to BCC transition is observed; arrows indicate the last BCC and first FCC structures in the simulation. At both given temperatures, the simulation exaggerates the miscibility of Cu and Fe significantly, especially on FCC lattices.

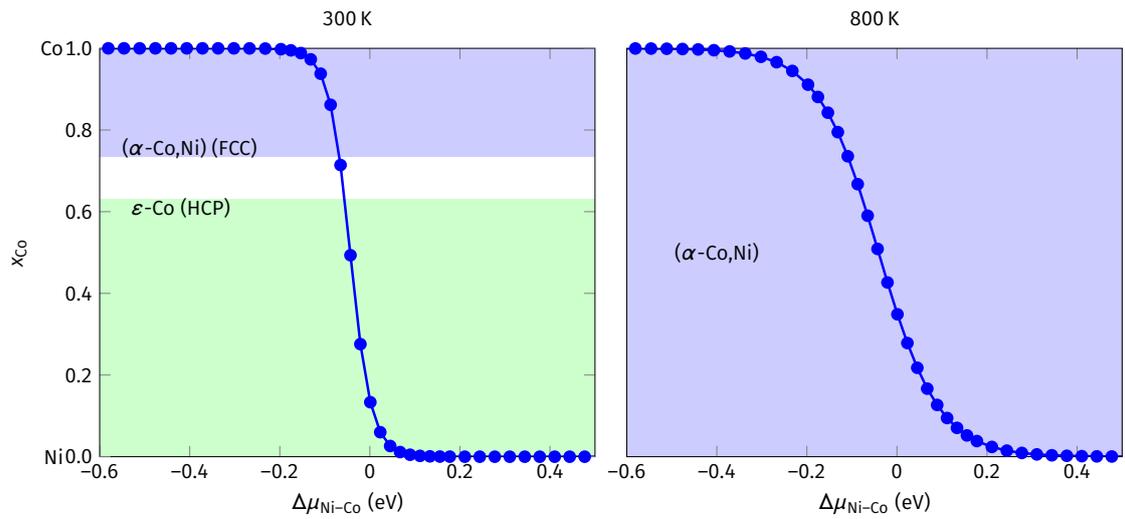

**Figure S.11:** Molar fraction of Co in the Ni–Co binary subsystem as a function of the chemical potential difference $\Delta\mu$ at 300 K and 800 K. Shaded areas indicate phases extracted and extrapolated from Nishizawa *et al.*[22] Due to constraints in the simulation setup, HCP phases do not occur. Therefore, an FCC solid solution is observed over the whole compositional range. The inability to transition from FCC to HCP renders us unable to observe the small miscibility gap at 300 K, similar to Fig. S.9. At 800 K the phase diagram is reproduced.

13

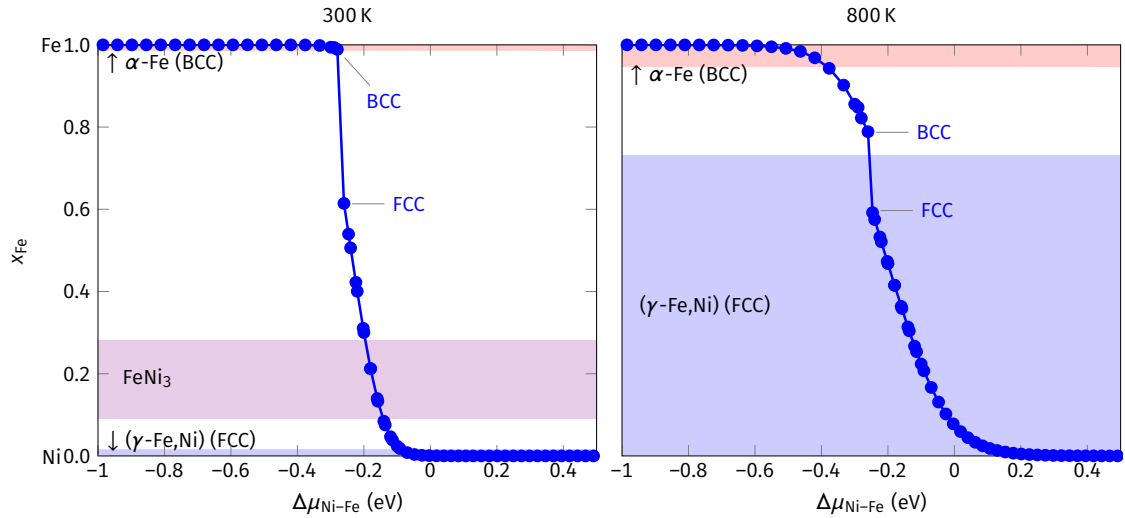

**Figure S.12:** Molar fraction of Fe in the Ni–Fe binary subsystem as a function of the chemical potential difference $\Delta\mu$ at 300 K and 800 K. Shaded areas indicate phases extracted and partially extrapolated from Swartzendruber *et al.*[12] The miscibility gap (indicated by an absence of data points) is underrepresented at 300 K. It should be noted, though, that $FeNi_3$ is an $L1_2$ phase, an ordered phase on an FCC lattice. Thus, while the potential does not capture the ordered phase, the extended FCC miscibility coincides with the $L1_2$ phase. At 800 K, the miscibility gap is shifted to lower concentrations. Arrows indicate the last BCC and first FCC structures observed in our simulations.

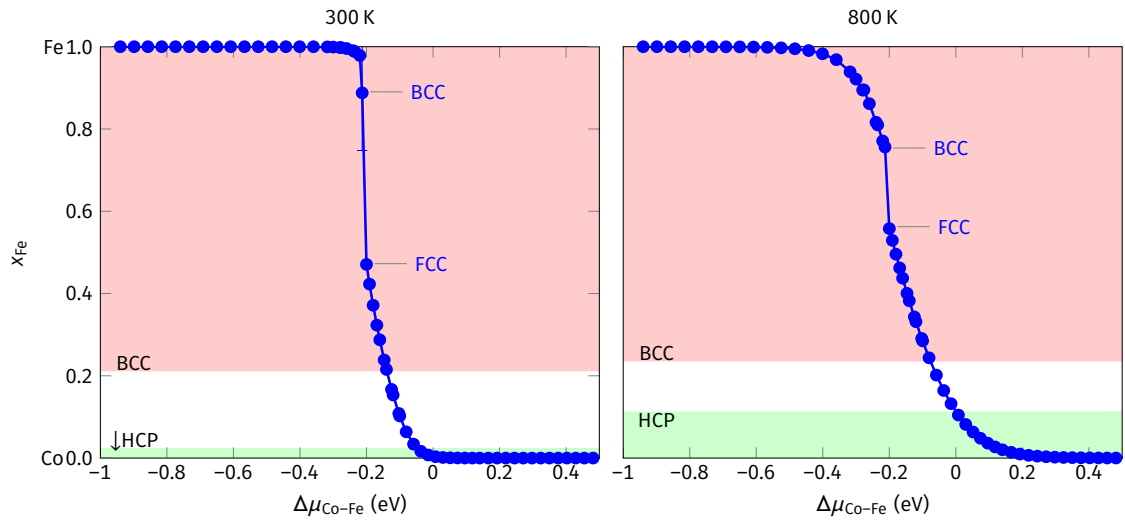

**Figure S.13:** Molar fraction of Fe in the Co–Fe binary subsystem as a function of the chemical potential difference $\Delta\mu$ at 300 K and 800 K. Shaded areas indicate phases extracted from Guillermet.[20] Due to constraints in the simulation setup, HCP phases do not occur. Therefore, an FCC to BCC transition is observed. The absence of HCP phases does not allow for a comparison of the miscibility gap (indicated by an absence of data points) with the phase diagram. At 800 K, the range of the miscibility gap is matched well, but it is strongly shifted to higher Fe concentrations. Arrows indicate the last BCC and first FCC structures observed in our simulations.


\end{document}